# Physical Effects in the Vicinity of the Ferroelectric-Antiferroelectric Interface

V.M. Ishchuk[1], V.L. Sobolev[2]

[1] Science & Technology Center "Reaktivelektron" of the National Academy of Sciences of Ukraine, Donetsk 83049, Ukraine
[2] Department of Physics, South Dakota School of Mines & Technology, Rapid City, SD 57701, USA

**Abstract**

Physical effects caused by the presence of interfaces between the domains of coexisting ferroelectric and antiferroelectric phases in solid solutions with small difference in free energies of the ferroelectric and antiferroelectric states are discussed in this review.

There exist a number of solid solutions in which the two-phase state of domains of the coexisting phases may occur under certain conditions. We present here results of investigations of some effects directly caused by the presence of the interfaces between the domains of the coexisting ferroelectric and antiferroelectric phases.

The phenomenological model describing the inhomogeneous state of coexisting domains of the ferroelectric and antiferroelectric phases is presented. Experimental studies of such inhomogeneous state are discussed using two systems of the $PbZr_{1-y}Ti_yO_3$-based solid solutions.

Detailed discussion of the process of local decomposition of solid solutions in the vicinity of ferroelectric-antiferroelectric interphase boundaries and formation of the mesoscopic system of segregates in the vicinity of these interphase boundaries and corresponding experimental results are presented.

Results on influence of the system of segregates on dielectric and piezoelectric properties, and also on a dielectric relaxation are presented and discussed. The effects caused by the application of a DC electric field are considered and corresponding experimental results are presented. The experimental results demonstrating the possibility of control of piezoelectric parameters by an external electric field in materials with the antiferroelectric to ferroelectric phase transition via the intermediate state of the coexisting domains of these phases are given.

# 1. Introduction

Inhomogeneous states of substances that contain domains of two coexisting phases, for instance, the substances undergoing the first order phase transitions are quite common in solid state physics. A stable state occurs within the hysteresis region in the diagram of phase states of the system in question. This hysteresis region is located between the boundaries of stability of each of the phases participating in the phase transition. The domains of both coexisting phases are present within a certain interval of thermodynamic parameters. Usually the presence of the interphase boundaries (the boundaries between the domains of coexisting phases) is considered to increase the free energy of the system with coexisting phases. In other words, the energy of the interphase boundary is positive (however, this fact has not been proven to be a general property of such systems so far).

Yet at the same time, a situation when this silently accepted rule seems to be violated is known at present. The example of such system is the type II superconductors placed in a magnetic field. In these superconductors the normal (resistive) and superconducting phases coexist within a field interval between the first and second critical fields [1, 2]. The inhomogeneous state in the type II superconductor is stabilized due to the negative surface energy of the boundary separating two interacting phases. However, such a phenomenon is observed only in magnetic field and is a consequence of the energy redistribution between the sample and the field: the energy of the inhomogeneous system decreases, whereas the density of the magnetic field energy rises over the whole space. The resulting energy of the system "superconductor-magnetic field" increases, and the inhomogeneous state of the superconductor is maintained due to the energy of the source of electric current creating this magnetic field. The evidence of such increase of the total energy is the fact that the superconductor regains its uniform state when the source of magnetic field is switched off.

The above example leads to a logical question: whether there exist other substances able to generate the field, which will lead to the effects observed in the type II superconductors (i.e. the inhomogeneous state consisting of domains of coexisting phases)? This question can be answered as follows. Such substances do exist – these are magnetic and ferroelectric materials at the temperatures below the Curie point. These substances possess the inhomogeneous states of domains of coexisting phases that are advantageous from the energy viewpoint in comparison with each of the uniform phases participating in the formation of this inhomogeneous state. Spontaneous appearance of the order parameter leads to advent of a field conjugate to this order parameter (scattering field). It should be noted that in ferroelectrics such fields do not penetrate the sample since they are shielded by the charges leaking on the surface; however, for our consideration this is not essential.

The inhomogeneous states of domains of coexisting phases are well known in substances with spontaneous magnetic polarization that undergo the first order phase transition from the antiferromagnetic phase to ferromagnetic phase (see for example [3]). In the substances with spontaneous electric-dipole polarization undergoing the first order phase transition from the ferroelectric phase to antiferroelectric phase the inhomogeneous state of domains of the coexisting ferroelectric and antiferroelectric phases is also possible. Among rather wide range of substances with electric-dipole polarization, there are compounds for which the two-phase state of domains of the coexisting phases may be realized under the certain conditions (certain values of thermodynamic parameters, i.e. temperature, field intensity, pressure etc.). Therefore, it seems very interesting to analyze the role of the field generated by the order parameter in formation of a inhomogeneous state of such systems characterized by the phase transition between the ferroelectric and antiferroelectric phases in the dipole ordered state.

The stable (from the viewpoint of energy) inhomogeneous two-phase state is possible in a dipole-ordered substance under two conditions [4, 5]. First, if the difference between the free energies of the antiferroelectric and ferroelectric phases is small within a wide interval of external parameters. Second, the interphase interaction between the domains of these two phases is taken into account. A uniform state of any of the ferroelectric or antiferroelectric phases will be metastable at these values of thermodynamic parameters.

The interdomain boundaries that appear in the process of formation of the above described inhomogeneous state of the substance play extremely important two-fold role. First, they make such inhomogeneous state more stable in comparison with the stability of the uniform ferroelectric and antiferroelectric states. Second, they form of a wide range of important properties caused by the presence of this type of interdomain boundaries in the substance. These interdomain boundaries are in fact the interfaces between two types of dipole-ordered phases, namely, the ferroelectric and antiferroelectric phases.

During a long history of investigations of ferroelectric materials the $PbZr_{1-y}Ti_yO_3$ (PZT) based solid solutions have been known as a system that possess a wide variety of phase transitions [6-9]. It so happened that the main attention of a large number of research groups has been devoted to studies of some particular compounds with the compositions close to the so called morphotropic phase boundary. The original "Ti-concentration–temperature" phase diagram for the PZT solid solutions [9] is given in Fig.1.1.

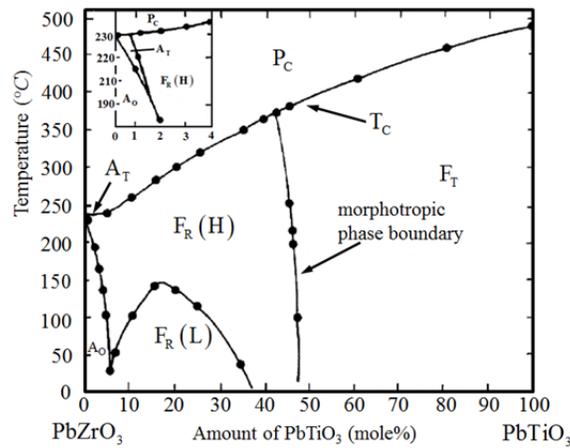

Fig 1.1. Phase diagram of the $Pb(Zr_{1-y}Ti_y)O_3$ series of solid solutions [9].

The morphotropic phase boundary is the nearly vertical boundary between the tetragonal and the rhombohedral phases and it is located in the vicinity of the 0.47 molar content of $PbTiO_3$ (see Fig.1.1). There are several excellent reviews devoted to studies of phenomena observed in ferroelectrics with compositions close to the morphotropic boundary [10, 11] where one can find most recent results as well as information about other ferroelectric systems with morphotropic phase boundary.

PZT-based solid solutions represent a perfect example of the systems in which the above-mentioned coexistence of the ferroelectric and antiferroelectric phases takes place. These two phases are present in the original phase diagram of PZT. The line separating ferroelectric and antiferroelectric phases is located in the region of small concentrations of $PbTiO_3$ (more details are in the insert in Fig.1.1). The substitution of the ions with smaller ionic radii for lead increases the energy stability of the antiferroelectric state with respect to the ferroelectric one. As a consequence the boundary separating the regions of the ferroelectric and antiferroelectric orderings in the original "Ti-content–temperature" phase diagram of PZT is shifted towards the higher concentrations of Ti [5, 12, 13-16]. The appearance of the phase diagram also changes. The diagrams now contain broad intermediate regions between the regions of the ferroelectric and antiferroelectric states. The example of modification of the part of the PZT phase diagram corresponding to the neighbouring regions of the antiferroelectric and ferroelectric phases in the process of the step by step substitution of lanthanum for lead giving the $(Pb_{1-3x/2}La_x)(Zr_{1-y}Ti_y)O_3$ (PLZT) system of solid solutions is presented in Fig.1.2 [5,13-16].

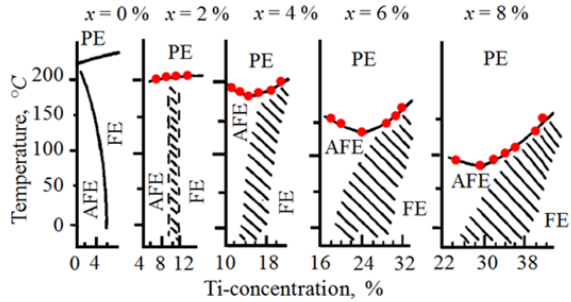

Fig.1.2. Modification of the original "Ti-composition–temperature" phase diagrams for the PZT system of solid solutions with step by step substitutions of La for Pb leading to formation of the PLZT system.

Here and further the notation, X/1-Y/Y, will be used for the composition of the solid solution derived from the PZT. The first number, X, corresponds to the percentage of the element substituting lead, whereas the second (1-Y) and the third Y numbers denote the content of zirconium and titanium, respectively. The border region in the phase diagrams in the Fig. 1.2 has no sharp boundaries. The region of the induced ferroelectric states appears in the "Ti-content–temperature" phase diagram after the samples have been subjected to the action of the external electric field. The appearance of the "Ti-content–temperature" phase diagrams after the action of the DC electric field for three series of PLZT solid solutions is presented in Fig. 1.3 [5, 13-16]. Fig. 1.3 [5, 13-16].

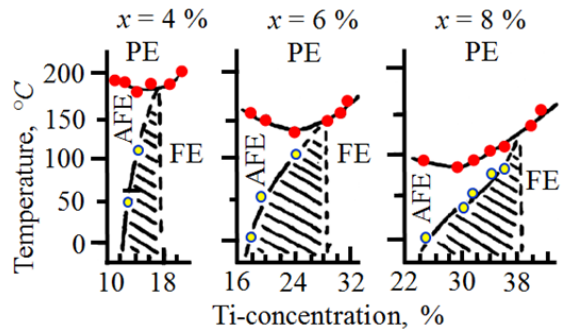

Fig.1.3. Appearance of the "Ti-content–temperature" phase diagrams after the action of a DC electric field for three series of the PLZT solid solutions.

Similar modifications of the phase diagrams have been observed in several PZT-based solid solutions: $Pb(Lu_{1-y}Ti_y)TiO_3$ $Pb_{1-x}(La_{1/2}Li_{1/2})_x(Zr_{1-y}Ti_y)O_3$ (PLLZT), $(Pb_{1-x}Sr_x)(Zr_{1-y}Ti_y)O_3$ (PSZT), and some other solid solutions.

This review presents the most interesting physical results that have been obtained in the PZT-based solid solutions that possess an interval of compositions within which the coexistence of the ferroelectric and antiferroelectric phases leads to the formation of a mesoscopic system of segregates formed in the vicinity of the interphase boundaries separating domains of the ferroelectric and antiferroelectric phases. Physical effect caused by the presence of this system of interphase boundaries are discussed using two particular systems of PZT-based solid solutions as an example.

The material of this review is organized in the following way. Second part of this paper contains results of investigations of the inhomogeneous state of coexisting ferroelectric and antiferroelectric phases in solid solutions with are characterized by close values of the free energies of the ferroelectric and antiferroelectric phases. The theoretical model for a phenomenological description of this inhomogeneous state is presented in the subsection 2.1. Results of experimental studies on visualization of the inhomogeneous state of the coexisting ferroelectric and antiferroelectric phases are given in the subsection 2.2.

The third part of this review is devoted to investigations of the process of the local decomposition of solid solutions in the vicinity of ferroelectric-antiferroelectric interphase boundaries and the formation of the inhomogeneous state of domain of these two phases.

The forth part describes some specific physical effects caused by the presence of the system of interfaces between the domains of the coexisting ferroelectric and antiferroelectric phases. The process of formation of the textured structures in the solid solution with the system of ferroelectric-antiferroelectric interfaces and its influence on some properties are described in the section 4.1. The phenomenon of a dielectric memory explained using the model for the solid solution with the inhomogeneous states of the coexisting domains of the ferroelectric and antiferroelectric phases is discussed in the section 4.2. Section 4.3 contains the description of peculiarities of dielectric properties and dielectric relaxation of solid solutions with coexisting ferroelectric and antiferroelectric phases. The effects caused by the application of a DC electric field are also discussed and corresponding experimental results are presented.

Section 4.4 contains results of theoretical consideration of the so-called intermediate state of materials with coexisting ferroelectric and antiferroelectric phases. This section also covers the experimental results that demonstrate the possibility of control of piezoelectric parameters by external electric field in materials with phase transitions via the intermediate state of coexisting domains of the ferroelectric and antiferroelectric phases.

Experimental results of studies of dependences of the light transmission coefficient on the wave length in materials with the negative refraction index created using the local decomposition of PZT-based solid solutions in the vicinity of interphase boundaries separating domains of the ferroelectric and antiferroelectric phases are presented in section 4.5

## 2. Inhomogeneous state of coexisting ferroelectric and antiferroelectric phases

### 2.1. *Theoretical model*

The thermodynamic potential for an infinite a two-phase system is presented as:

$$\varphi = \xi_1 \varphi_1 + \xi_2 \varphi_2 + W_{int} \tag{2.1}$$

where $\varphi_1$ and $\varphi_2$ are the thermodynamic potentials of the phases; $\xi_1$ and $\xi_2$ are the volume shares occupied by each of the phases; and $W_{int}$ describes the interaction between the domains of these phases.

The energy of the phase interaction can be represented as follows [4]:

$$W_{\text{int}} = \frac{1}{2}\left\{\xi_2 \sum_{i,j}\left[E_{\eta_{1,i}}(x_2,y_2,z_2)\eta_{2,j}(x_2,y_2,z_2)\right] + \xi_1 \sum_{i,j}\left[E_{\eta_{2,j}}(x_1,y_1,z_1)\eta_{1,i}(x_1,y_1,z_1)\right]\right\} \qquad (2.2)$$

Here $\eta_{1,i}$ and $\eta_{2,j}$ are the order parameters of the first and second phases, respectively ($i$ and $j$ are the numbers of the order parameters of these phases); $E_{\eta_{1,i}}(x_2,y_2,z_2)$ is the field, induced in the domains of the second phase by nonzero value of the order parameter of the first phase; $E_{\eta_{2,j}}(x_1,y_1,z_1)$ is the field in the domains of the first phase, initiated by nonzero value of the order parameter of the second phase. Based on the physical nature of the order parameters (or the nature of the coexisting phases), it is evident that not all the products entering expression (2.2) are nonzero. This is a consequence of the symmetry of order parameters and fields influencing the order parameters. The only nonzero terms in the expression (2.2) are those that transformed according to the representations containing the identity representation.

The dependences of the fields $E_{\eta_{1,i}}(x_2,y_2,z_2)$ and $E_{\eta_{2,j}}(x_1,y_1,z_1)$ on the order parameters generating them have a complex form. To determine these dependences one must take into consideration the physical nature of the order parameters, the particular shape of the domains, as well as the spatial distribution of the order parameters over the volume of the domain. This problem must be solved using a self-consistent procedure, taking into account the situation when the fields change the spatial dependences of order parameters, which, in their turn, define these fields.

Such a problem does not seem to be solved exactly at present and, therefore various approximations are used to find the solution. We shall restrict ourselves by the expansion into a series of the powers of $\eta_{\alpha,i}$:

$$E_{\eta_{1,i}}(x_2,y_2,z_2) = \xi_1 C_{1,i}(x_2,y_2,z_2)\eta_{1,i} + \cdots \qquad (2.3a)$$

$$E_{\eta_{2,i}}(x_1,y_1,z_1) = \xi_2 D_{2,i}(x_1,y_1,z_1)\eta_{2,i} + \cdots \qquad (2.3b)$$

Note that the considered fields are rather long-range and smoothly vary in space.

In the case when external fields are present, they may be taken into account in (2.1) in a usual way.

Thus, the study of behavior of the system of domains of interacting coexisting phases is reduced to the investigation of the thermodynamic potential (2.1) with $W_{int}$ in the form (2.2), along with the expressions (2.3) under the condition $\sum_\alpha \xi_\alpha = 1$ ($\alpha$ = 1, 2). The next step is a conventional procedure of minimization for the nonequilibrium thermodynamic potential

$$\varphi_\lambda = \varphi - \lambda\left(\sum_\alpha \xi_\beta - 1\right) \qquad (2.4)$$

where $\lambda$ is the indeterminate Lagrangian multiplier. This procedure leads to the system of equations for the determination of the equilibrium values of the order parameters:

$$\xi_\alpha(\partial\varphi_\alpha/\partial\eta_{\alpha,i} + E_{\eta_{\alpha',i}}) = 0, \qquad (\xi_\alpha \neq 0), \qquad (2.5)$$

$$\varphi_\alpha + \eta_{\alpha,i} E_{\eta_{\alpha',i}} = \lambda = const, \quad \alpha, \alpha' = 1, 2. \tag{2.6}$$

As seen from Eq. (2.6), the condition for existence of the thermodynamically equilibrium structure of domains of the coexisting phases is the equality of their thermodynamic potentials, taking into account the fields (2.3), but not the equality of the "bare" thermodynamic potentials. It also follows from Eq. (2.6) that there is a peculiarity of the considered multiphase structure connected with the spatial dependence of the coefficients in (2.3). This peculiarity is observed inside the region separating the domains of the coexisting phases which is wider than the usual domain boundary. The fields $E_{\eta_{\alpha,i}}$ are spatially dependent, and their intensity decreases while moving inside the domains of the other phase. Therefore, in these boundary regions the thermodynamic potentials differ from those, which characterize the inner regions of domains. This fact testifies that the phase state inside the region separating the domains will not be similar to the inner regions of the adjacent domains. It is obvious that the phase state of this region separating the domains of different phases is transient between the states of the adjacent domains.

For a particular example of the ferroelectric and antiferroelectric phases, the simplest density of the nonequilibrium thermodynamic potential has the form [17, 18, 4]:

$$\varphi = \frac{\alpha_1}{2} P^2 + \frac{\alpha_2}{4} P^4 + \frac{\beta_1}{2} \eta^2 + \frac{\beta_2}{4} \eta^4 + \cdots + \frac{A}{2} P^2 \eta^2 \tag{2.7}$$

Here $\alpha_2$, $\beta_2$ and $A$ are positive values ($A > \sqrt{\alpha_2 \beta_2}$), $\alpha_1$ and $\beta_1$ may change their sign at the temperatures $T_{c,f}$ and $T_{c,af}$, respectively:

$$\alpha_1 = \alpha_0 (T - T_{c,f}); \quad \beta_1 = \beta_0 (T - T_{c,af}). \tag{2.7a}$$

Considering the thermodynamic potential in the form (2.7), one has to take into account that the antiferroelectric phase transition is a structural transition into the state with the order parameter $\eta$ (see [17], for example), and the latter interacts with the polarization $P$, which is the order parameter of the ferroelectric state [19]. If $T_{c,f} > T_{c,af}$, the expression (2.7) describes the phase transition between the paraelectric and ferroelectric states at varying temperature, if $T_{c,f} < T_{c,af}$, (2.7) describes the paraelectric-antiferroelectric transition.

For definiteness let us consider the case when the ferroelectric phase (phase 1) is more stable in energy than the antiferroelectric one ($T_{c,f} > T_{c,af}$) and investigate the peculiarities of behavior of the system.

To write down the thermodynamic potential for each of the phases let us refer to Fig.2.1 in which the schematic equipotential lines of the nonequilibrium thermodynamic potential (2.7) as the $P$ - $\eta$ plot are presented. This schematic map of equipotential lines corresponds to the temperature interval within which there exist local minima, corresponding to possible low-temperatures states of the system. The minimum corresponding to the ferroelectric phase has the coordinates ($P_{1,0}$, 0), where $P_{1,0}^2 = -(\alpha_1/\alpha_2)$. The equilibrium energy value of this state is $\varphi_{1,0} = -(\alpha_1^2 / 4\alpha_2)$. The minimum corresponding to the antiferroelectric state has the coordinates (0, $\eta_{2,0}$), $\eta_{2,0}^2 = -(\beta_1/\beta_2)$. The equilibrium value of the energy of this state is $\varphi_{2,0} = -(\beta_1^2 / 4\beta_2)$.

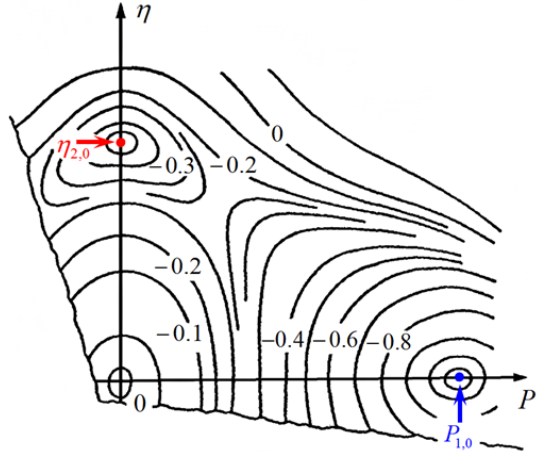

Fig.2.1. A schematic map of equipotential lines for the thermodynamic potential (2.7) in a temperature interval $T < T_{c,af} < T_{c,f}$

Now the nonequilibrium potentials for each of the phases may be written (up to the quadratic terms) as:

$$\varphi_1 = \varphi_{1,0} + U_1\left(P_1 - P_{1,0}\right)^2 + V_1\eta_1^2 \qquad (2.8a)$$

$$\varphi_2 = \varphi_{2,0} + U_2\left(\eta_2 - \eta_{2,0}\right)^2 + V_2 P_2^2. \qquad (2.8b)$$

The coefficients $U_i$ and $V_i$ may be expressed in terms of the coefficients in expansion (2.7). The most important for our consideration is that they are positive.

In accordance with (2.2) and (2.3), taking into account the restrictions imposed by the order parameters symmetry (see the remark accompanying expressions (2.2)), the interphase interaction energy has the following form:

$$W_{int} = \xi_2 P_2 E_{P_1} + \xi_1 \eta_1 E_{\eta_2} = \xi_1\xi_2 C_1 P_{1,0} P_2 + \xi_1\xi_2 D_2 \eta_{2,0} \eta_1 \qquad (2.9)$$

Now, according to (2.1), (2.8) and (2.9), the density of the nonequilibrium thermodynamic potential of the two-phase system can be written as:

$$\varphi = \xi_1\varphi_{1,0} + U_1\left(P_1 - P_{1,0}\right)^2 + V_1\eta_1^2 + \xi_2\varphi_{2,0} + U_2\left(\eta_2 - \eta_{2,0}\right)^2 + V_2 P_2^2 + \xi_1\xi_2 C_1 P_{1,0} P_2 + \xi_1\xi_2 D_2 \eta_{2,0}\eta_1, \qquad \xi_1 + \xi_2 = 1. \qquad (2.10)$$

A conventional procedure of minimization with respect to $P_1$, $\eta_1$, $P_2$ and $\eta_2$ gives the equilibrium values of these parameters:

$$P_1 = P_{1,0}, \qquad \eta_1 = -\frac{1}{2}V_1\xi_2 D_2 \eta_{2,0}, \qquad \eta_2 = \eta_{2,0}, \qquad P_2 = -\frac{1}{2}V_2\xi_1 C_1 P_{1,0}. \qquad (2.11)$$

Substitution of (2.11) into (2.10) gives the density of equilibrium energy for the two-phase system under consideration:

$$\tilde{\varphi} = \xi_1\left(\varphi_{1,0} - \frac{\xi_2^2}{4V_1}D_2^2\eta_{2,0}^2\right) + \xi_2\left(\varphi_{2,0} - \frac{\xi_1^2}{4V_2}C_1^2P_{1,0}^2\right). \tag{2.12}$$

Comparison of expressions (2.12) and (2.8) shows, that the phase interaction in the form (2.9) lowers the energy of the system. As the final result, this two-phase state may acquire larger energy stability than a single phase state, if the difference of $(\tilde{\varphi} - \varphi_{1,0})$ is negative. The latter condition is fulfilled in the vicinity of the ferroelectric-antiferroelectric stability line. In this case $\varphi_{1,0} \cong \varphi_{2,0}$, $\xi_1 = \xi_2 = 1/2$, and the absolute energy advantage due to the formation of the inhomogeneous two-phase state will be equal to:

$$\Delta W = -\frac{1}{32}\left(\frac{D_2^2}{V_1}\eta_{2,0}^2 + \frac{C_1^2}{V_2}P_{1,0}^2\right). \tag{2.13}$$

As seen from (2.3), the fields $E_1$ and $E_2$ and, consequently, the coefficients $C_1$ and $D_2$, are the functions of the coordinates. They decrease as the distance from the interphase boundary increases. Thus, the long-range interaction fields favour the stabilization of inhomogeneous state to the greatest degree. The stability of the inhomogeneous state is also raised by low values of the parameters $V_1$, $V_2$, $U_1$ and $U_2$.

In the considered case of the coexisting ferroelectric and antiferroelectric phases the long-range fields have electric (dipole) and elastic nature. Since in each real crystal there exist free charges, crystal lattice defects (e.g. dislocations), the radius of action of these fields is not infinite, and corresponds to the distance equal to several tens of lattice parameters. Therefore the energy advantage of the two-phase state (13) is ensured by the regions separating domains of the phases in the sample's volume.

For a more general case than considered here (given by the equation (2.9)), the expression for the interphase interaction energy has the form:

$$W_{int} = \xi_1\xi_2 C_1 P_1 P_2 + \xi_1\xi_2 D_2 \eta_1 \eta_2. \tag{2.14}$$

This case was considered in [4, 20], and it was found that the energy stabilising the inhomogeneous state was:

$$\Delta W = -\frac{1}{32}\left(\frac{D_2^2}{V_1}\eta_{2,0}^2 + \frac{C_1^2}{V_2}P_{1,0}^2\right) - \frac{1}{512}\left(\frac{C_1^4}{U_1 V_2^2}P_{1,0}^2 + \frac{D_2^4}{U_2 V_1^2}\eta_{2,0}^2\right) \tag{2.15}$$

As it can be seen, the inhomogeneous state becomes even more stable when more general case is considered.

The above consideration demonstrates that the interaction between the coexisting ferroelectric and antiferroelectric phases (the phases described by different order parameters) actually stabilizes the inhomogeneous state of domains of these coexisting phases. The negative value of $\Delta W$ in (2.13) (and in more general case (2.15)) demonstrates in fact that the considered interphase boundary possesses a negative surface energy. But such a negative energy will lead to the division of the sample volume into unlimited number of very small domains (thus ensuring the largest area of the interphase boundaries and, consequently, the largest energy advantage).

The presence of different phases and their interaction have been confirmed experimentally by an X-ray diffraction and a transmission electron microscopy. However, it should be emphasized that the notions of phase and phase state are defined and introduced for rather large finite volume of substance (theoretically, for the volume which tends to infinity). This means that division of a finite volume of the substance into domains of the coexisting phases cannot last indefinitely. This process will have to end at a certain stage. The latter is defined by the least size of the phase domains at which the phases still exist. At further reduction of the domain size the phases disappear and one cannot speak of their existence as well as an interphase interaction and the energy of interphase boundaries.

A simplified consideration yet reflecting the main physical result is presented here. For example, specific effects associated with the conditions of continuity for elastic medium at the interphase boundary have not been taken into account. The interphase domain wall separates the domains with the ferroelectric and antiferroelectric states. Elementary crystal cells of the ferroelectric and antiferroelectric phases have different size. In connection with the above-said two, distinctive features should be emphasized. First, it has been demonstrated [21] that at the conjugation of crystal planes of the phases with close crystal structures (this is the situation takes place when the ferroelectric and antiferroelectric phases coexist), which differ in interplane distances, intervals between possible dislocations are of order of several tens of *nm*. Second, the TEM studies of the coexisting domains of the ferroelectric and antiferroelectric phases showed that linear sizes of these domains are also of order of several tens of nanometers [13, 14], and no dislocations at the interphase domain boundaries were found.

This indicates that the crossing of the interphase domain wall (from one phase to the other) is accompanied with continuous conjugation of the crystal planes (free of breaks and dislocations). Such a coherent structure of the interphase domain wall leads to an increase of the elastic energy. This increase is the more essential the larger is the difference in the configuration volumes of the ferroelectric and antiferroelectric phases.

The above-discussed effect gives a positive contribution into the surface energy density of the boundaries separating ordinary domains in ferroelectrics [22, 6]. This elastic energy weakens the condition of existence of the inhomogeneous state. The $Pb_{0.90}(Li_{1/2}La_{1/2})_{0.10}(Zr_{1-y}Ti_y)O_3$ system of solid solutions was used to demonstrate how the change in the interplane distances of the ferroelectric and antiferroelectric phases (achieved by the variation of the Ti-content of solid solutions) influences the formation of the inhomogeneous state of domains of coexisting phases. In the process of change of the Ti-content of solid solution the mechanism responsible for the formation of domains of coexisting phases changes from the mechanism providing the elastic blocking (caused by the striction) in the process of formation of domains of the metastable phase to the mechanism providing the inhomogeneous state of coexisting domains of the ferroelectric and antiferroelectric phases.

Thus, the analysis of the system described by the thermodynamic potential (2.1) with the interphase interaction (2.2) shows that the stable state of this system may be inhomogeneous (see Eq. (2.6) and the commentary to it). For the substances characterized by ferroelectric-antiferroelectric phase transition and described by the potential (2.7) it is shown that this inhomogeneous state may be considered as the state of the coexisting domains of the ferroelectric and antiferroelectric phases with a negative energy of the interphase boundary.

The possibility of existence of the inhomogeneous states with the negative energy of the interphase boundary leads to two significant conclusions. The first conclusion is that it gives the possibility of a simple, an elegant and self-consistent physical explanation of different aspects of known phenomena that have not been adequately treated so far. The second conclusion concerns the prediction (and subsequent experimental verification) of earlier unknown phenomena in different physical processes.

The above considerations were dealing with the stability of the two phase state at temperatures below the Curie point. Some physically correct conclusions regarding the peculiarities of behavior of the considered system may be also made for temperatures close to $T_{c,f}$ ($T_{c,f} > T_{c,af}$) and above it. The thermodynamic potential of the system has only one minimum within the interval of temperatures close to the paraelectric transition. Therefore, the substance described by such potential is ferroelectric, and the appearance of domains of the antiferroelectric phase at temperatures near $T_{c,f}$ is possible only in the form of fluctuations. These domains emerging in the substance's volume interact with the ferroelectric matrix. Such interaction changes the density of thermodynamic potential in the volume of the sample within which it takes place. The density of the potential describing the ferroelectric state remains unchanged in the other parts of the volume.

Thus, for the volume, in which the interphase interaction takes place, the density of the nonequilibrium thermodynamic potential may be written as [4, 20]:

$$\varphi = \frac{\alpha_1}{2}P_1^2 + \frac{\alpha_2}{4}P_1^4 + \frac{\beta_1}{2}\eta_1^2 + \frac{A}{2}P_1^2\eta_1^2 + \frac{m}{2}P_2^2 + \frac{n}{2}\eta_2^2 + CP_1P_2 + D\eta_1\eta_2. \qquad (2.16)$$

In this expression the first four terms correspond to the density of the nonequilibrium thermodynamic potential of the ferroelectric phase for $T > T_{c,af}$ ($\beta_1 > 0$, so we assume that $\beta_2 = 0$), the next two terms correspond to the density of the nonequilibrium potential of fluctuational domains of the antiferroelectric phase ($m > 0$, $n > 0$), and the last two terms represent the interphase interaction.

The minimization of (2.16) with respect to $P_2$ and $\eta_2$ yields:

$$P_2 = -(C/m)P_1; \qquad \eta_2 = -(D/n)\eta_1. \qquad (2.17)$$

Substituting the obtained expression into (2.16) we have:

$$\varphi = \frac{1}{2}\left(\alpha_1 - \frac{C^2}{m}\right)P_1^2 + \frac{\alpha_2}{4}P_1^4 + \frac{1}{2}\left(\beta_1 - \frac{D^2}{n}\right)\eta_1^2 + \frac{A}{2}P_1^2\eta_1^2. \qquad (2.18)$$

It follows from Eq. (2.18) that the interphase interaction leads to the renormalization and increase of the Curie point (see the coefficient at $P_1^2$). Moreover, under certain conditions this interaction may also stabilize the antiferroelectric state in some part of the sample's volume (see the coefficient at $\eta_1^2$). This means that there is a possibility of existence of the two-phase (ferroelectric + antiferroelectric) domains at temperatures exceeding $T_{c,f}$.

It is easy to find the distribution of the Curie temperatures through the crystal volume (that is the dependence of the Curie temperature $T_c$ on spatial coordinates). It is sufficient to use the nonequilibrium thermodynamic potential (2.7) for the system with two close depth minima of the thermodynamic potential corresponding to two dipole-ordered states and with the temperature dependences of the coefficients $\alpha_1$ and $\beta_1$ taken in the form of Eq. (2.7a).

As it was shown earlier in the region of space where the interphase interaction takes place the energy change defined by Eq. (2.13) must be added to Eq. (2.7). In this expression the order parameters $\eta_{2,0}$ and $P_{1,0}$ correspond to the part of the crystal free of interphase interaction ($\eta$ and $P$ featured in Eq. (2.7)) that is these are the order parameters determined by the thermodynamic potential (2.7). Thus, the coefficients of the expansion (2.7) are renormalized as follows

$$\alpha_1 = \alpha_0 \left[ T - \left( T_{c,f} + \frac{1}{32} \frac{C_1^2(x,y,z)}{V_2} \right) \right] \qquad \beta_1 = \beta_0 \left[ T - \left( T_{c,af} + \frac{1}{32} \frac{D_2^2(x,y,z)}{V_1} \right) \right]. \qquad (2.19)$$

As may be inferred from the above expression the local temperatures of the phase transition into the paraelectric phase increase and the spatial dependence of the Curie temperature appears due to the spatial dependences of $C_1(x, y, z)$ and $D_2(x, y, z)$.

It has to be reminded that the coefficients $V_1$ and $V_2$ in Eq. (2.18) are determined by the coefficients of the thermodynamic potential (2.7) and are positive as it was mentioned above. That is why the spatial distribution of the Curies temperatures is not only determined by the interphase interaction but by the original properties of the system as well.

For example, in the systems with the anharmonicity of the crystal's elastic potential the flattering of the minima of potential (2.7) leads to a decrease of the $V_1$ and $V_2$ coefficients. The anharmonicity of elastic crystal potential also leads to the widening of the distribution of temperatures of the phase transition into completely disordered state. It also follows that the domains in which the state of the coexisting spatially separated ferroelectric and antiferroelectric phases is realized (in the first approximation) become stable in the temperature region of original paraelectric state.

As one can see form Eqs. (2.18) and (2.19) the local regions of crystal located in the vicinity of boundaries separating domains with the ferroelectric and antiferroelectric orderings are characterized by lower free energy than the main volume of the crystal. At the temperatures above the Curie temperature such domains of ordered phases initially emerge as fluctuations and then they get stabilized by the interphase interaction. At approaching the Curie temperature during cooling from higher temperatures the domains with $P \neq 0$ and $\eta \neq 0$ are the first to emerge in the paraelectric matrix and then in the vicinity $T_c$ they acquire the features of the two-phase domains. The temperature of emergence of the first nuclei of the dipole-ordered state may exceed the Curie temperature by $100°C$ and even more (see for example [5, Ch. 6]).

Our simplified approach does not take into account the effects connected with the presence of elastic stresses caused by the phase coexistence and the interphase boundaries. Therefore some additional remarks have to be done. It is a general knowledge that the configuration volume increases in the course of the paraelectric-ferroelectric phase transition, whereas at the transition into the antiferroelectric state the configuration volume decreases. Therefore the existence of the domains of ferroelectric and antiferroelectric phases separately at temperatures exceeding the Curie points is accompanied by the appearance of essential elastic stresses, and is not advantageous from the energy point of view.

However, the increase of the energy does not take place when the complex two-phase (ferroelectric + antiferroelectric) domains appear. In this case the configuration volume does not vary for the two-phase domain as a whole. The presence of domains of such structure allows to remove the elastic stresses and to raise the dipole disordering temperature due to the interaction (2.9) as much as possible (i.e. to decrease the free energy to the utmost). The ratio of the phase volumes in the two-phase domain which is present in the paraelectric matrix of the crystal is defined by both the relative stability of the ferroelectric and antiferroelectric phases (i.e. by the relative difference of their free energies) and the changes in the elastic energy, which may be brought about by each of the phases. A characteristic property of such a domain is that it should easily match the varying external conditions: electric field, pressure (including those of the paraelectric matrix). An insignificant polarization of this domain, almost complete absence of elastic stresses and the absence of distortions in the paraelectric matrix, where the domain emerges, provide its existence at temperatures exceeding both $T_{c,f}$ and $T_{c,af}$. For simplicity the latter phenomenon may be thought of as a complex two-phase nucleation. It is realized in the form of a spatial domains present in the paraelectric matrix of the crystal.

**2.2.** *Experiments on visualization of the inhomogeneous state*

First experimental confirmations of the above-presented results were obtained in [13, 14]. Direct observation of coexisting domains of the ferroelectric and antiferroelectric phases in the samples of the 7/65/35 PLZT was carried out by transmission electron microscopy and later in the samples of the 8/65/35 PLZT [24]. Position of these solid solutions in the "Ti-content-temperature" phase diagram corresponds to the hysteresis area separating the regions of stable ferroelectric and antiferroelectric states. The last circumstance means that these phases possess practically equal phase stability having close values of their free energy. Antiferroelectric state is the main phase state of these solid solution at room temperatures (it possesses a deeper minimum of the free energy) while the ferroelectric state is metastable.

A coarse-grained ceramics with 8 to 10 $\mu m$ size of crystallites was used for observation of the domain structure. The spalls of certain crystallites (produced along different crystallographic planes) with a thickness of not more than 0.2 $\mu m$ were chosen for the observation of domain structure using a JEM-200 transmission electron microscope. These spalls were essentially lamella-type crystals of PLZT cut out along different crystallographic directions. Such method of sample preparation allowed to avoid clamping due to the surrounding crystallites which is always present after sample preparation by thinning of ceramic samples as well as to avoid physical effects caused by this clamping. Transmission electron microscopy images of the structure of the 7/65/35 PLZT for two crystallite spalls are given in Fig. 2.2.

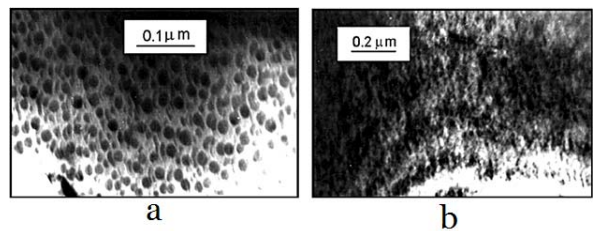

Fig. 2.2. Transmission electron microscopy bright-field images for the 7/65/35 PLZT.
*a* – for (110) the foil plane, *b* – for (120) the foil plane [13, 14].

The bright-field image for the sample with the (110) spalling plane (the incident electron beam was along the [110] direction) is given in Fig.2.2a. The two-phase structure in the form of ellipsoidal dark regions in a lighter shade antiferroelectric matrix is clearly visible at this orientation of the crystallite. The domain diameter is $(2 - 4) \cdot 10^{-6}$ *cm*, and their density is of the order of $10^{11}$ $cm^{-2}$. The opacity of these regions is caused by the additional scattering of electron beam by the depolarization field of the ferroelectric domain. In spite of the difference in interplane spacing of these two phases the dislocations are not present at the interphase boundaries. Specific manifestations of distortions of the crystal matrix that can be produced by the elastic stresses caused by the metastable phase domains are also absent in this image.

Electron diffraction images of the same crystallite manifest the system of reflections that are unique for the perovskite crystal structure of the PLZT solid solutions without the presence of any reflections corresponding to irrelevant inclusions.

The two-phase domain structure of crystallite oriented in such a way that the direction of the incident electron beam is the [120] crystal direction is presented in Fig. 2.2b. In this picture the same domain structure is under the oblique incidence of electron beam which gives the domain structure image form the side. The filamentary structure of domains ingrown through the crystallite thickness is clearly visible. Similar images have been obtained for the other spalls of crystallites in the case when their [110] directions were misaligned with respect to the electron beam.

Analysis of images with different orientations of crystal planes with respect to the incident electron beam showed that the domains of metastable phase have a cylindrical shape with the axes oriented along the [110] direction. These domains are grown through the thickness of the crystallite if their spalling plane was (110). Such shape of domains of the metastable ferroelectric phase occurs in thin crystals only. In the bulk samples the shape of these domains is ellipsoid of revolution close to a sphere [25-27].

There exist a number of recently obtained transmission electron microscopy images of the PLZT solid solutions. Usually these images are obtained on samples with different preparation prehistory. As a rule the authors of a particular publication do not always pay close attention to this matter. But the sample prehistory is specifically the feature that determines the phase composition of the solid solutions with compositions belonging to the hysteresis region in the phase diagram. With rear exception the authors of these publications see the structures shown in Fig. 2.2b. As an example we want to mention papers [28, 29] in which the room temperature studies of the structure of annealed (and left to age for some time) samples and the structure of samples subjected to AC electric field with intensity larger than the coercive field of the studied solid solutions were carried out. The author's approach was that they did not search for coexisting domains in particular and because of this they did not keep watch for specific orientation of their samples with respect to the electron beam.

Studies in a paper [30] following the publication with direct identification of the system of coexisting domains contains results of the transmission electron microscopy obtained on samples obtained by thinning of the ceramic specimens by mechanical grinding and polishing, chemical etching and the ion-beam etching at the final stage. In the process of such preparation of samples the clamping of the crystallite chosen for observation by the neighboring crystallites always takes place. Thus, the imaging of the structure takes place under the presence of the lateral mechanical stress (in the plane of observation). This lateral stress may be compressive or tensile. The image of the domain structure observed in [30] is shown in Fig. 2.3. As one can see the system of cylindrical domains has transformed into the system of stripe domains. Similar transformation of one domain system into another is possible in the system of cylindrical magnetic domains.

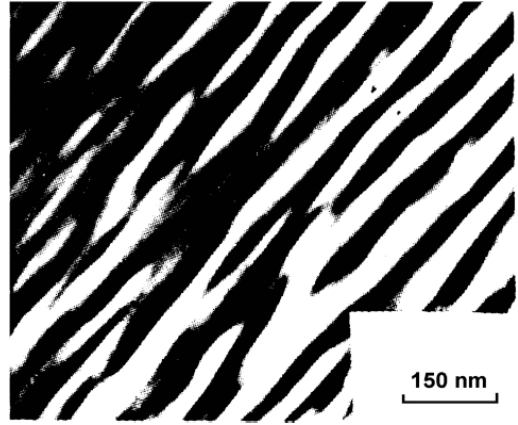

Fig. 2.3. Aligned macrodomain structure in PLZT [30].

In relation to the above said it is interesting to follow up the changes of domain structure existing at nanoscale during the variation of the relative stability of ferroelectric and antiferroelectric phases (for example, due to the solid solution's content variation primarily due to the change of the La-content). A transmission electron microscopy of the X/65/35 and X/60/40 (located close to X/65/35 in the phase diagram of PLZT) series of the PLZT solid solutions was carried out in [31]. The crystal structure of the ferroelectric phase (when its stability slightly exceeds the stability of the antiferroelectric phase) is a perovskite structure with the tetragonal distortions only in the X/60/40 series of solid solutions whereas the crystal structure of the ferroelectric phase in the X/65/35 series of solid solution manifests the rhombohedral distortions. Increase of the lanthanum concentration in specified solid solutions leads to the transition from the region of ferroelectric states ($X \leq 5$ for X/65/35 series) into the region of antiferroelectric states ($X \geq 8.5$) through the hysteresis region ($6.0 < X < 8.0$) in the phase diagram.

Bright-field images of the microstructure of X/65/35 series of PLZT with the clearly traceable transition of the domain structure with changes of the lanthanum concentration are given in Fig. 2.4 [31]. Superfine domain structure with the nanometer scale exists in solid solutions with $X \geq \sim 8$ (Fig. 4.2a). Sizes of domains of the antiferroelectric phase are of the order of $20 - 30$ *nm*, which is in agreement with results of [13, 14, 24] obtained earlier. Increase of both the share of antiferroelectric phase and the size of its domains takes place when the lanthanum concentration decreases (that is the stability of the antiferroelectric phase increases) is clearly visible in the image of the 7/65/36 PLZT (Fig.2.4b). In some cases the ferroelectric domains combine into the system of stripe domains in the antiferroelectric matrix. Authors of [30] also noted that the recharging of samples taking place during studies in electron microscope did not practically result in the domain motion which in turn confirmed the increase of stability of the ferroelectric state one more time.

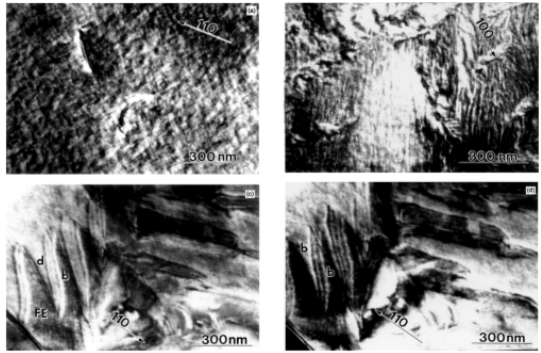

Fig. 2.4. Room temperature transmission electron microscopy images of various X/65/35 PLZT compositions [31] revealing:
(*a*) mottled contrast typical of relaxor ferroelectrics, imaged in dark field, X = 8;
(*b*) domains with striation-like morphology, imaged in bright field, X = 7;
(*c*) ferroelectric domains (labelled FE) and asymmetric δ fringes (labelled b, d), X = 5;
(*d*) the same grain as in (c), imaged in dark field, revealing symmetric δ fringes (labelled *b*).

The fully established system of ferroelectric domains on micrometer scale is observed in the 5/65/35 PLZT (Fig.2.4*c,* 2.4*d*). If to refer to the phase diagram of the X/65/35 PLZT system of solid solutions it is easy to see that the 5/65/35 PLZT is located in the region of spontaneous ferroelectricity at room temperature but its location is close to the hysteresis region of the ferroelectric-antiferroelectric phase transition. According to the above-considered model (see section 2.1) the interphase interaction lowers the energy of the two-phase ferroelectric-antiferroelectric domain with respect to the energy of the homogeneous ferroelectric state. This circumstance is manifested in the transmission electron microscopy images.

"Ordinary" ferroelectric domains (such domain is marked as FE in Fig. 24c) possess clearly expressed boundaries, the twinning planes are mainly {110}. Domain ordering in these planes is "head-to tail" and they can be considered as 109° domains [32]. Twinning planes {100} separating 71° domains are essentially less common. Authors of [30] note (even though it is not clearly seen in the images) the presence of spatial regions (with contrast of the type shown in Fig.2.4a) between the "ordinary" ferroelectric domains. This contrast is characteristic for the structure of nanometer scale domains of the coexisting ferroelectric and antiferroelectric phases. Precisely in these regions of crystallites the mechanical stresses are concentrated. These stresses increase the stability of the antiferroelectric state with respect to the ferroelectric state and stabilize the nonhomogeneous state of domains of coexisting phases in the crystal with composition actually corresponding to the region of spontaneous ferroelectricity in the phase diagram.

The so-called δ-fringe appears in their vicinity of the ("ordinary") domain boundaries inclined with respect to the plane of the foil (it is marked as *b* and *d* in the Fig. 2.4c and 2.4d). This δ-fringe is visible more clearly in the dark-field images. The structure of wedges that limit this δ-fringe represents the line structure of nanoscale domains. The reason for appearance of such structure in the images is some (weak) misorientation of crystallographic planes in the neighboring nanodomains and as a consequence a misorientation of the scattering vectors of electrons. The most probable reason for appearance of such domain structure is the increase of stability of the two-phase ferroelectric-antiferroelectric state and nucleation of the system of coexisting domains of these phases. Such formations are observed in the regions of concentration of elastic stresses. In these cases the side (in the plane of the image) stresses lead to the transformation of the system of cylindrical domains into the system of stripe domains. In connection with the above-said it ought to be noted that such system of stripe domains was first observed in [33] in 1974 which seems like a long ago at present. Most likely the majority of researches are not aware of this particular study.

Let us now dwell on the X/60/40 system of solid solutions [31]. Transmission electron microscopy images of the samples with different concentrations of lanthanum, X, are shown in Fig.2.5. Dissimilarity of this series of solid solution from the above-considered X/65/35 series is that the boundary of spontaneous ferroelectric state at room temperature is located at slightly higher concentration of lanthanum ( at X close but slightly higher than 8) and the crystal structure of this state possesses the tetragonal type of distortions.

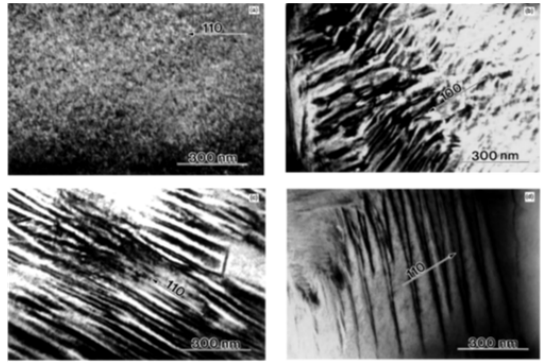

Fig. 2.5. Room temperature images for various X/60/40 compositions of the PLZT solid solutions [31] revealing: (*a*) mottled contrast, bright field image, X = 10; (*b*) coarser domains formed near grain boundaries and mottled contrast at grain centers, bright field image, X = 9; (*c*) ferroelectric domains with stripe-like morphology, bright field image, X = 8; (*d*) wedge-shaped domains, bright field image, X = 8.

In general the physical picture observed in this series of solid solutions is similar to the one observed in the X/65/35 PLZT. The microstructures of the 10/60/40 and 9/60/40 are similar and represent the coexisting domains of the ferroelectric and antiferroelectric phases with sizes of the order of 20–30 *nm*. Enlargement of the metastable ferroelectric domains and their merging is noticeable during the transition from the first system of solid solutions to the second one. One can observe small-scale stipe domains in the vicinity of domain boundaries. The appearance of domains of such form is caused by mechanical stresses. It is quite analogous to the situation taking place in the 8/65/35 and 7/65/35 PLZT solid solutions. The stripe domain structure emerges with the further decrease of lanthanum content in solid solution (that is at the increase of stability of the ferroelectric state). This stripe domain structure is different from the stripe domain structure of nanometer scale existed in the 5/65/35 PLZT solid solution (Fig.2.4c and 2.4d). One can observe the remains of the small scale structure that exists in the 10/60/40 PLZT (Fig. 2.5a)

The primary domain structure in the 8/60/40 solid solution contains ordinary 90°-domains with polarization axes in neighboring domains directed perpendicularly to one another. According to [31] these very domains form the domain structure presented in Fig.2.5c. Wedge-shape domains are also present in the 8/60/40 solid solution. The presence of such domains is responsible for appearance of the {100} type of twin planes. As it was noted in [34] such domains appear in $PbZrO_3$ and PZT solid solutions with small titanium concentrations during the ferroelectric-antiferroelectric phase transition caused by the change of temperature. It is again the evidence of the fact that the energy stability of the ferroelectric and antiferroelectric states differs only slightly.

Thus, the behavior of the microstructure of the PLZT solid solutions at variation of relative stability of the ferroelectric and antiferroelectric phases has been considered above. This variation has been achieved by changing the lanthanum concentration at a constant ratio of concentration of titanium and zirconium. All presented results are in complete correspondence with the model of inhomogeneous state of coexisting domains of the ferroelectric and antiferroelectric phases given above in section 2.1.

In conclusion let us discuss the behavior of the same microstructure during the variation of relative stability of the dipole-ordered phases due to the change of the ratio of concentrations of titanium and zirconium that is along the horizontal cross section of the phase diagram of PLZT corresponding to the lanthanum concentration $X = 8$. Corresponding transmission electron microscopy images [31] are given in Fig.2.6 (the images presented earlier are also given here for reader's convenience). It has been noted above that at room temperature the 8/60/40 is located in the spontaneous ferroelectricity region of the phase diagram. The domain structure of this solid solution corresponds (taking into account all notes and comments given above) to the "ordinary ferroelectric" (Fig.2.6.c). The antiferroelectric state stability rises with the increase of zirconium concentration and the antiferroelectric phase becomes more stable than ferroelectric in the 8/65/35 solid solution but still close to the boundary of the region of stability of the antiferroelectric phase. This means that there are two free energy minima and based on the model of interacting phases a consequence the existence of these minima leads to the presence of nanoscale domains of the coexisting ferroelectric and antiferroelectric phases. As one can see in Fig.2.6b the domains of the indicated-above phases coexist in this solid solution with approximate equality of the shares of both phases.

Further stabilization of the antiferroelectric phase happens with further increase in the concentration of zirconium. Even though the free energy of the 8/70/30 PLZT solid solution still has two minima but the minimum of the antiferroelectric phase is much deeper than the minimum of the ferroelectric phase. As a consequence the share of antiferroelectric phase has to prevail in the solid solution with this composition. This conclusion is amply demonstrated by the image in Fig. 2.6a.

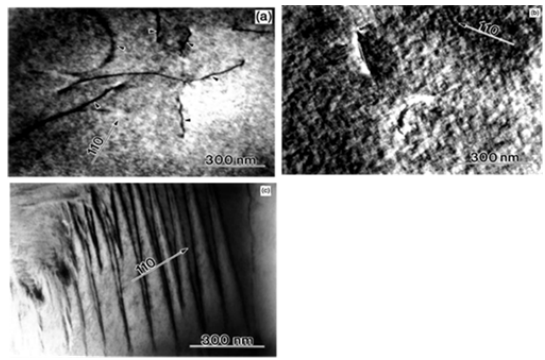

Fig. 2.6. Bright-field images of domain structure in the PLZT solid solutions with lanthanum content X = 8 [31]
a − 8/70/30, b − 8/65/35, c − 8/60/40.

Properties of the PLZT solid solutions with compositions from the region of induced states and peculiarities of their behavior (especially under the action of electric field since external pressure is used very seldom due to the cumbersome experimental procedure) are widely discussed in the literature and there is no doubt that the induced phase is a ferroelectric phase. All back-and-forth discussions are about the nature of the state before the field was applied. Nowadays when the complete set of the phase diagrams for these solid solutions have been obtained (see, for example, [5, 35]) the antiferroelectric nature of this initial state is already evident. Peculiarities of this state are determined by the small difference in energies of two dipole-ordered states. Due to such small difference in energies the metastable ferroelectric phase exists in antiferroelectric matrix of crystallites. The two-phase state turns out to be more stable than single phase one.

All above-presented results of transmission electron microscopy studies fully corroborate the two-phase character of states of the PLZT solid solutions with compositions form the hysteresis region of the phase diagram. In addition to what was said above we would like to turn to results of [30]. The authors of this article studied the domains of stable phase in solid solutions from the hysteresis region of the phase diagram of PLZT. The system of the $\left(h+\frac{1}{2}, k+\frac{1}{2}, l+\frac{1}{2}\right)$-type reflections is present in electron diffraction images besides of the base Bragg reflections at room temperature. The intensity of reflections rises slightly with decrease of temperature. The dark-field image of the part of crystallite obtained using one of these fine structure reflections is given in Fig. 2.7a. The antiphase boundaries (they are marked by arrows) separating the antiferroelectric domains (as it will be shown slightly further) are observed.

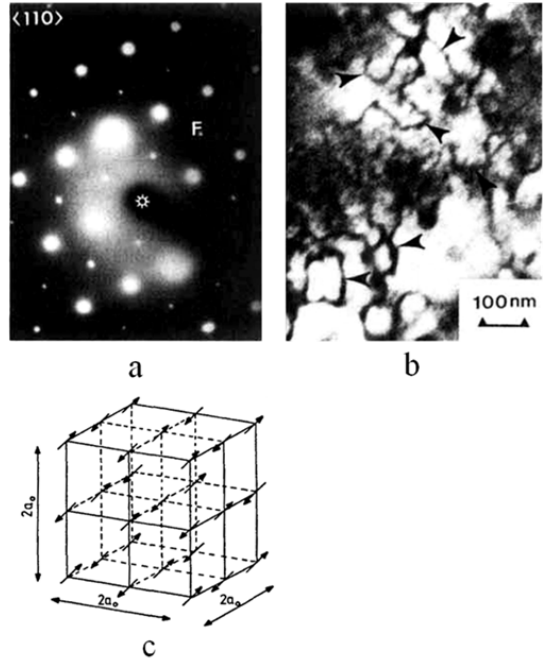

Fig. 2.7. $\langle 110 \rangle$-zone axis diffraction pattern (*a*), dark field image showing domains associated with *F*-spot in 8.2/70/30 PLZT, at T = − 40 ºC (*b*) and antiparallel ordering in the PLZT antiferroelectric structure (*c*) [30].

Usually the system of the fine structure reflections of the $\left(h+\frac{1}{2}, k+\frac{1}{2}, l+\frac{1}{2}\right)$-type is exhibited in the following several cases:
– during ordering in the B-sublattice [36]
– during the rotation of octahedrons about the main diagonal of the cubic of perovskite crystal lattice [37]

– in the case of antiparallel dipole ordering (the antiphase displacements of the lead ions along the $M_5'$ mode [38]) in antiferroelectric phase of lead zirconate and in lead zirconate based solid solutions, for example PZT.

The first case listed above does not suggest the existence of the temperature dependence of the fine structure reflections in the temperature interval close to room temperature. As for two other cases of appearance of the $\left(h+\frac{1}{2}, k+\frac{1}{2}, l+\frac{1}{2}\right)$-type reflections they accord fully with the antiferroelectric state of the solid solutions doped with small amount of lanthanum ions (with the rotations of the oxygen octahedrons about the [111] axis). This situation was analyzed in [30] and the authors suggested the arrangement of the ordered dipoles (given in Fig. 2.7 with antiparallel location of dipoles in the lattice sites in the [111] plane of the crystal lattice of PLZT).

The authors of the study [39] entertained slightly different point of view. They related the appearance of the $\left(h+\frac{1}{2}, k+\frac{1}{2}, l+\frac{1}{2}\right)$ reflections with the phase transition into the state analogous to the low temperature rhombohedral state of PLZT. While at the same time they noted that the same $\left(h+\frac{1}{2}, k+\frac{1}{2}, l+\frac{1}{2}\right)$ reflections were present in the solid solutions with small concentrations of lanthanum. They stated that the appearance of these reflections in these solid solutions had an unambiguous interpretation as the presence of the antiferroelectric state because the softening of the $\Gamma_{25}(R)$-mode plays essential role in stabilization of the antiferroelectric state in the lead zirconate and in PZT-based solid solutions.

A different method was used for studies of the coexisting phases in the X/65/35 series of the PLZT solid solutions in [40]. The small scale structure of domains of coexisting phases was studies by means of a piezoelectric microscope. Hot pressed ceramic samples of the 9.75/65/35 PLZT with the thickness of the order of 0.3 – 0.5 *mm* polished to the optic quality were used in these experiments. Experimental method permits to register the specific patterns created at the sample surface by the domains in the bulk and thus register the "emergence of domains" at the sample surface. Unlike the transmission electron microscopy this method is "less direct," however, it allows extracting effects caused by the interaction of the constant electric field allowing at the same time to keep track of the material's behavior at the nanoscale level of resolution. To the best knowledge of the authors of present review nobody could achieve this level of resolution using transmission electron microscopy thorough to 2013 and even the transmission electron microscopy method itself does not allow doing it (at least as of now). To this must be added that the procedure of sample preparation for this studies is much simpler.

As it follows form the phase diagram of PLZT [5, 13-16] the basic state of the 9.75/65/35 solid solution at room temperature is the antiferroelectric state with an admixture of the ferroelectric phase as an "impurity". This very structure has been observed by authors of [5] at room temperature. It worth to discuss results obtained in [40] in greater detail because such type of studies has not been adequately covered for the time being (in comparison, for example, with transmission electron microscopy studies). Some apparatus features are the following. The tip with the radius of the tip apex slightly less than 10 *nm* was used. The tip was retained against a sample's surface by the spring with rigidity of 35 *N/m*. The visualization of domains was assured by application of 4 *V* of AC electric potential with a frequency of 5 kHz (far from the resonance frequency of the cantilever).

The topography and piezoelectric response (at two magnifications) from the surface of the ceramic sample of the 75/65/35 PLZT is shown in Fig.2.8. As one can see at this magnification the structure of grains is inhomogeneous consisting in polar (piezoelectrically active) and nonpolar domains. One can see in Fig. 2.8 that the polar axes in nanodomains from neighboring crystallites are oriented in different spatial directions, which are pointed out by the distribution of their response (but this does not mean that they are oriented along different crystallographic axes). Domain sizes are determined in the interval from 30 to 50 nm which is in agreement with the sizes (although slightly bigger) obtained using transmission electron microscopy (Fig.2.2 and Fig.2.3). The dependence of the dipole moment of the polar domain on its area exposed at the polished sample's surface is determined as $P(S)=P_0 S^{D/2}$ [40] with $D \approx 1.55$. This value is close to 1.50 that has to be observed at completely random distribution of polar directions [41, 42]. Some discrepancy can be due to the depolarization fields generated by polar domains.

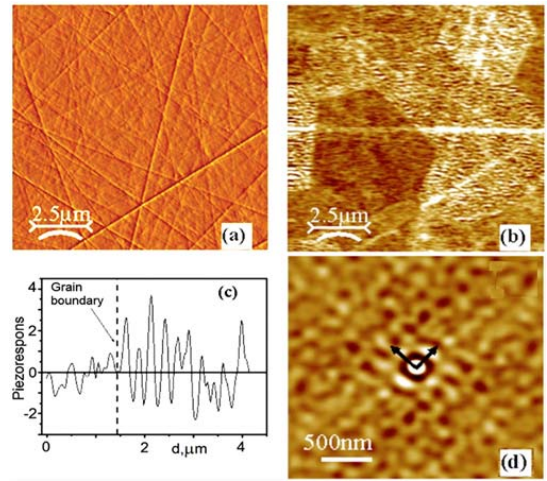

Fig. 2.8. Typical topography (a) and piezoresponse (b) and (d) images of the 9.75/65/35 PLZT ceramics. Cross section of piezoelectric response image (c) shows the change of piezoelectric response in adjacent grains [40].

When DC electric field is applied to the nonpolar antiferroelectric domain an induction of the ferroelectric state takes place (which is demonstrated in Fig. 2.9) and the piezoelectric hysteresis loop is observed when the electric field intensity varies. In the same image one can clearly see the appearance of contrast in the region where induction of the ferroelectric phase took place. The reverse transition takes place when the electric field is removed however this process is the long-term one. The contrast that has been preserved during 7 hours after the field was switched off is visible in Fig. 2.9. The mechanism of such long-term relaxation consists in proximity of the free energies of the ferroelectric and antiferroelectric phases at room temperature. It will be discussed below in detail.

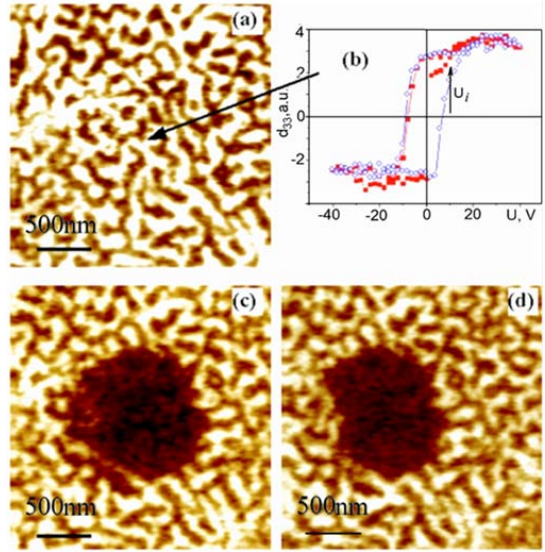

Fig.2.9. Initial piezoelectric response image (*a*), local piezoelectric response hysteresis loop (*b*) (arrow indicates position where the loops were measured), piezoelectric response image just after the application of a bias field (*c*), and piezoelectric response image relaxation during 7 h (*d*) [40].

## 3. Kinetics of the local decomposition of solid solution in the vicinity of the ferroelectric-antiferroelectric interface and creation of the inhomogeneous state

In the case when domains of the coexisting ferroelectric and antiferroelectric phases are present in the sample these domains are separated by the interphase domain walls (the PZT-based solid solutions are the most striking example of such systems). Crystal lattices inside these domains have the elementary crystal cells with different sizes. As it was shown in Part 2, there are no dislocations present at the interphase walls. This signifies that crossing of the interphase domain walls (from one phase into the other) is accompanied with the continuous conjugation of atomic planes (free of breaks and dislocations). Such a coherent interphase domain wall structure leads to increase of the elastic energy. This increase is the more essential the larger is the difference in the configuration volumes of the ferroelectric and antiferroelectric phases.

In the substances where the ferroelectric and antiferroelectric states are present, equivalent crystallographic sites are occupied by the ions that differ either in their sizes or in their charges or in both charges and sizes. In a single-phase state (or inside the domains of each of the coexisting phases) each of the ions, forming the crystal lattice, is not subjected to the action of forces (more correctly, the sum of all the forces acting on each ion is zero) in the absence of external factors. Entirely opposite situation takes place for the ions located near the "bare" interphase domain wall. The balance of forces affecting each of the ions is upset. "Large" ions are pushed out into those domains which have a larger configuration volume and, consequently, a larger distance between crystal planes. "Small" ions are pushed out into the domains with a smaller configuration volume and smaller interplanar distances. Such process is accompanied with both a decrease of the elastic energy concentrated along the "bare" interphase domain wall and an increase of the energy bound up with the segregation of the substance. The considered process of ion segregation will be completed when the newly-formed structure of the interphase domain walls provides the minimum of energy. Such a "clothed" interphase domain wall will be further called the real interphase domain wall or simply the interphase domain wall.

Such mesoscopic segregations in have been observed in experiments [43, 44] for the first time. The mechanism of their formation was not clear at that time. However the dependence of parameters of these segregation on the solid solutions composition (that is on the location of the solid solution in the phase diagrams), and the influence of DC electric field on the formation of segregations were studied [44].

Here we consider the kinetics of segregation in the process of the local decomposition of solid solution in the vicinity of interphase domain walls separating domains of the coexisting ferroelectric and antiferroelectric phases and the formation of the inhomogeneous state. The investigation was performed on two series of the PZT-based solid solutions based solid solutions that possess domains of the ferroelectric and antiferroelectric phases. To extend the region of solid solution compositions where the phase coexistence occurs, the isovalent complex $(La_{0.5}Li_{0.5})^{2+}$ or $La^{3+}$ ions were used as substitutions for lead ions in PZT [4, 5, 35, 45].

The "Ti-content–temperature" phase diagrams of the PZT-based system of solid solutions, obtained at sequential substitution of lead ions by the $(La_{0.5}Li_{0.5})^{2+}$ complex (PLLZT) and by the $La^{3+}$ ions (PLZT), are presented in Fig.3.1a and Fig.3.1b, and demonstrate similar character.

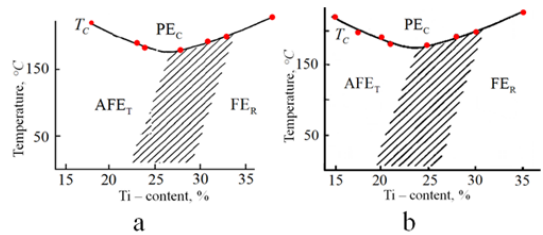

Fig.3.1. "Ti-content-temperature" phase diagrams for the 6/100-Y/Y PLZT solid solutions (a) and for the 15/100-Y/Y PLLZT solid solutions (b).

As it was already mentioned in Introduction the borderland between pure phases (shown by dashes in Fig.3.1), which separates the regions of the ferroelectric and antiferroelectric ordering, is shifted toward higher concentrations of titanium in the solid solutions as the content of the $(La_{0.5}Li_{0.5})^{2+}$ complex and the $La^{3+}$ ions substituting lead increases. This transition region is the hysteresis region for the ferroelectric-antiferroelectric transformation. The procedure of obtaining of the Ti-content – temperature phase diagrams for PLZT and the results of their investigation are described in [4, 35, 46, 47] in greater detail. For PLLZT this procedure is considered in [23, 48, 49].

The "Ti-content–temperature" phase diagrams for all the series with the content of lanthanum higher than 4% for the case of PLZT solid solutions and with the content of $(La_{0.5}Li_{0.5})^{2+}$ higher than 10% for PLLZT solid solutions are equivalent from the viewpoint of physics. Therefore, the investigations in the scope of the present work were carried out on PLZT with 6% content of lanthanum, i.e. on the 6/100-Y/Y PLZT series and on PLLZT with 15 % of $(La_{0.5} Li_{0.5})^{2+}$, i.e. on the 15/100-Y/Y PLLZT series.

The results of structural investigations of PLLZT solid solutions with zirconium-titanium contents varying within in a wide concentration range are reported in [23]. In the frames of the present work, the most significant for our consideration is fact that for the solid solutions located within the limits of the dashed region of the Ti-content–temperature phase diagram the FE and AFE phase domains coexist in the samples' bulk (Fig.3.2).

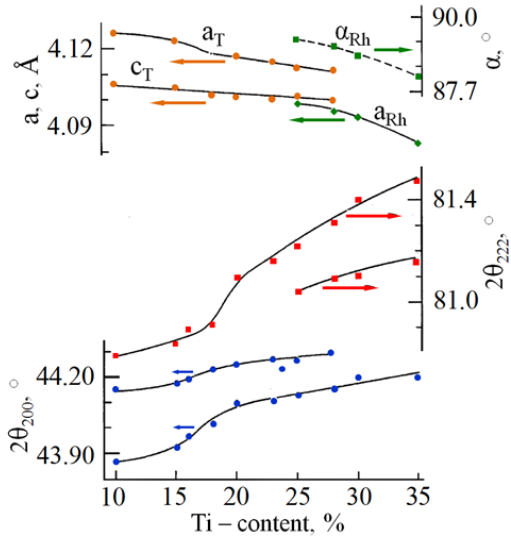

Fig.3.2a. Dependences of the crystal cell parameters on Ti content for the 6/100-Y/Y PLZT series of solid solutions (at the top). Dependences of the positions (the $2\theta_{200}$ and $2\theta_{222}$ angles) of the peaks for the (200) and (222) X-ray lines on Ti content (at the bottom). The parameters $a_T$ and $c_T$ correspond to the tetragonal antiferroelectric phase and the crystal cell parameter $a_{Rh}$ and the rhombohedral angle, $\alpha_{Rh}$, are for the rhombohedral ferroelectric phase.

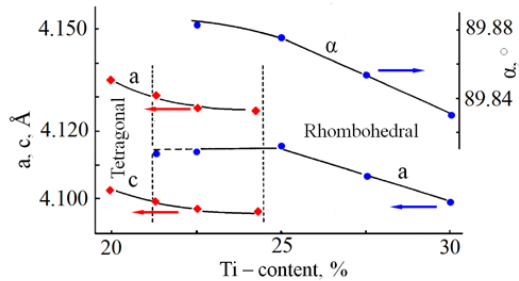

Fig.3.2b. Dependences of the crystal cell parameters on Ti content for the 15/100-Y/Y PLLZT series of solid solutions. Here a and c are the lattice parameters of the tetragonal antiferroelectric phase and the cell parameter a and the rhombohedral angle, $\alpha$, correspond to the rhombohedral ferroelectric phase.

The data presented above demonstrates that domains of the ferroelectric and antiferroelectric phases coexist within the dashed regions in Fig.3.1 in both the 15/100-Y/Y PLLZT series and the 6/100-Y/Y PLZT series of solid solutions. The coexistence of domains of the ferroelectric and antiferroelectric phases in the volume of samples with compositions from the borderland in the Ti-content–temperature phase diagrams has also been confirmed by X-ray and transmission electron microscopy studies [13, 14, 23, 24, 47].

Investigations of the crystal structure were carried out using a SIEMENS D-500 powder diffractometer with Ge monochromator ($CuK_{\alpha 1}$ radiation with a wave length of 1.54056 Å) and BRAUN gas filled position sensitive detector. The accuracy of the 2θ angle measurements was 0.01 degree. The pulse accumulation time for the 200, 222, and 400 lines was 1*s*, 2 *s*, and 3 *s* respectively. X-ray diffraction confirmed that the samples were single-phase with perovskite crystal structure. Ceramic samples used in the majority of our studies had grain sizes in the interval from 5 to 7 *μm*.

The decomposition of solid solution is manifested in the appearance of weak diffusion lines (halos 1 and 2 in Fig.3.3, pattern 2) in the X-ray diffraction patterns near the basic diffraction lines characteristic for the perovskite crystal structure of the solid solutions investigated [44, 50, 51]. Therefore, while studying the kinetics of long-duration relaxation and the formation of mesoscopic structures of segregations, there arises the problem of record of X-ray diffraction patterns in a wide range of the angles $\theta$ during a short period of time. These studies have been carried out by the method that was used in studies of the atomic ordering processes in the oxide ferrite substances [52, 53] as well as at examining clustered structures in the PZT-based solid solutions [43, 44, 24]. This method consists in a registration of scattered X-rays from samples placed in a Debye X-ray chamber with subsequent photometry of X-ray diffraction patterns (the Debye–Scherrer method). In our investigations we used a chamber with a diameter of 57 mm. The registration was carried out by the method of "plane section" with $CoK_\alpha$ radiation filtered by a layer of vanadium oxide selective absorber. The layer thickness was chosen experimentally. The process of registration lasted for 20 *min* (10 *min* for each of the two positions of the sample plane symmetric with respect to the incident X-ray beam). The angular speed of a flat sample was 1*rpm*.

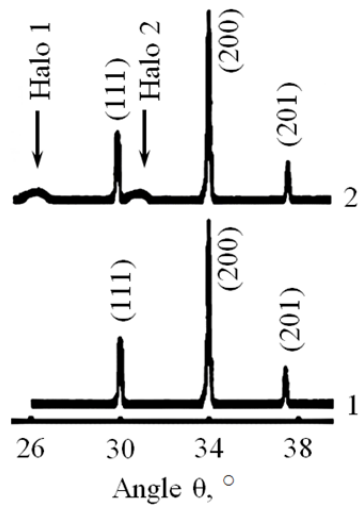

Fig.3.3. Debye–Scherrer X-ray diffraction patterns of the 6/73/27 PLZT solid solution obtained at the room temperature right after the quenching (1) and after ageing during 30 days (2).

The measurements were carried out according to the following scheme. The samples were annealed at 650°C during 22 hours. After such an annealing they were quenched to room temperature. Then the samples were aged at room temperature during the period of time, $\tau$, and afterwards X-ray diffraction patterns were obtained.

The solid solutions belonging to the hysteresis region in the Ti-content–temperature phase diagram, namely, the 15/77/23 PLLZT and the 6/73/27 PLZT were chosen for experiments. In these solid solutions the domains of ferroelectric phase (with rhombohedral type of distortions of the elementary perovskite cell) and antiferroelectric phase (with tetragonal type of distortions of the elementary perovskite cell) coexist in the samples' bulk at room temperature.

The X-ray patterns of the PLZT solid solutions obtained at $600°C$ on the annealed samples contained only strong singlet X-ray lines caused by the coherent scattering from crystal planes of the perovskite cubic lattice. It is important to note that the ferroelectric Curie temperatures of the solid solutions under investigation are well below the annealing temperature, but well above the ageing temperature (room temperature).

After the high-temperature X-ray studies the samples were quenched to room temperature and then left to age at 22°C during the time interval, τ. At the end of each time interval τ the X-ray studies were carried out by the Debye–Scherrer method. Right after quenching (τ ≈ 0) the X-ray patterns contained only strong singlet diffraction lines as in the high-temperature case (Fig. 3.3 (pattern 1). The structure of X-ray patterns becomes more complicated during ageing. A splitting of the singlet lines takes place. Broadened diffuse lines (halos) with a significantly lower scattering intensity appear in addition to the diffraction lines (Fig. 3, pattern 2). These new lines are caused by the incoherent scattering from chaotically oriented segregates at the interphase ferroelectric–antiferroelectric boundaries [24, 27, 28]. We studied the behavior of the halos located inside two intervals of angles, namely, $\theta = 25°–27°$ (Halo 1) and $\theta = 29°–32°$ (Halo 2). The intensity, location, and shape of the halo changed with time. The shape and location of the diffraction lines characterizing the crystal structure of the solid solution under investigation also changed with time.

The dependences of the elementary cell volume (the calculations were made based on the position of the (200) X-ray diffraction peak in the pseudo-cubic approximation) on the ageing time for the above mentioned solid solutions are shown in Fig.3.4. In the first stage of the process of ageing (of about 15–20 h for PLZT and about 3 h for PLLZT), the elementary cell volume of the perovskite crystalline structure of solid solution increases. At longer ageing times, the volume decreases. As one can see, the change of the volume virtually follows an exponential behavior for both the first and the second stage of ageing.

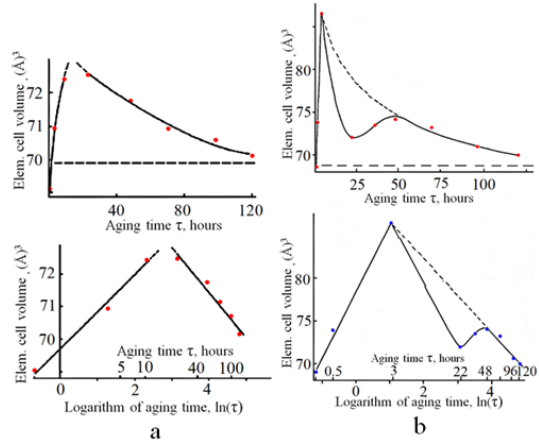

Fig.3.4. Dependences of the elementary cell volume on the ageing time for the 6/73/27 PLZT solid solution (a) and for the 15/77/23 PLLZT solid solution (b).

The peculiarity of dependence of the elementary cell volume on aging time V(τ) that takes place at τ ≈ 22 h will be discussed later.

The shape, intensity and position of the halos as functions of ageing time are presented in Fig.3.5a, Fig.3.5b, Fig.3.6a, and Fig.3.6b. The following peculiarities of these dependences attract attention. The shape, angular position and intensity of diffuse scattering lines changes during the ageing process. It should be mentioned that there is a clear correlation between the change of positions of the X-ray diffraction lines (Fig.3.7) and the change of location of the diffuse scattering lines. The dependence of intensity of the diffuse lines on the ageing time reaches saturation in approximately 25 h.

Changes in the shape and position of X-ray diffraction lines during the ageing process are given in Fig.3.8. We analyzed the behavior of the (111) and (200) X-ray lines, which are the most typical for the perovskite structure of the solid solutions under investigation. The (111) X-ray line is a singlet in the case of tetragonal lattice distortions, and it is a doublet in the case of rhombohedral distortions. On the contrary, the (200) X-ray line is a doublet in the first case and a singlet in the second case.

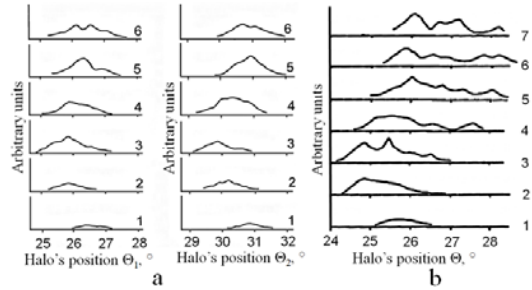

Fig.3.5. Changes of the profiles and positions of two halos in the process of ageing after the quenching of the 6/73/27 PLZT solid solution (a). Ageing time (hours): 1—0.5, 2—3.5, 3—23, 4—48, 5—72, 6—120.
Changes of the profile and position of the halo in the process of ageing after the quenching of the 15/77/23 PLLZT solid solution (b). Ageing time (hours): 1—0.3, 2 – 0.5, 3—3.0, 4—22, 5—48, 6—72, 7—120.

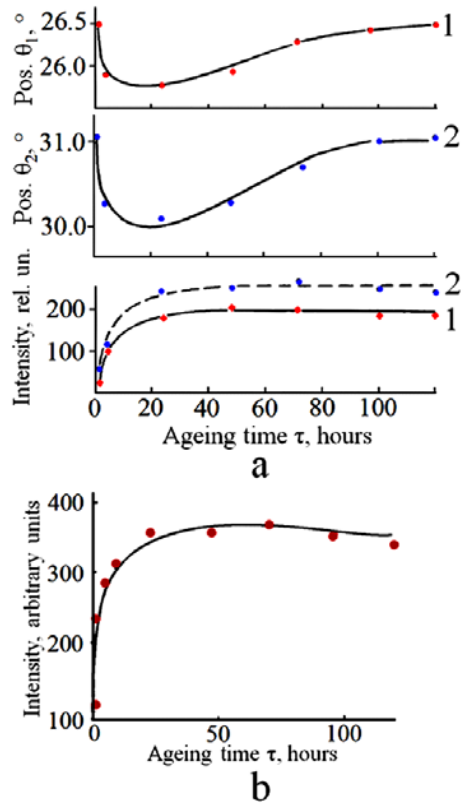

Fig.3.6. a) Positions and intensities of two diffuse lines vs. ageing time for the 6/73/27 PLZT solid solution (halo 1– curve 1 and halo 2 – curve 2).
b) Ageing time dependence of the position of diffuse line for the 15/77/23 PLLZT solid solution.

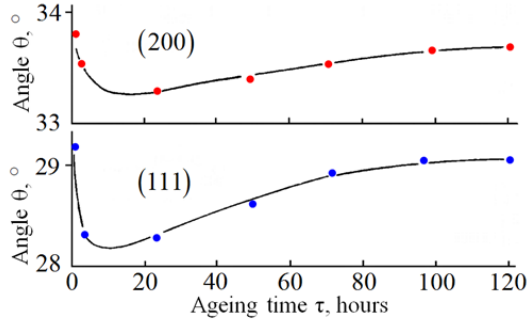

Fig.3.7. Positions of the (111) and (200) X-ray diffraction lines vs. ageing time for the 6/73/27 PLZT solid solution.

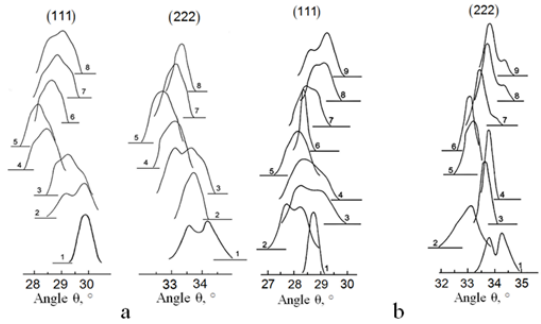

Fig.3.8. (a) Variations in the shape and position of the (111) and (200) X-ray diffraction lines in the process of the samples' ageing after quenching of the 6/73/27 PLZT solid solution. Ageing time (hours): 1—0.25, 2—0.5, 3—3.5, 4—23, 5—48, 6—72, 7—96, 8—120. (b) Variations in the shape and position of the (111) and (200) x-ray diffraction lines in the process of the samples' ageing after quenching of the 15/77/23 PLLZT solid solution. Ageing time (hours): 1—0.3, 2 – 0.5, 3—3.0, 4—22, 5 - 48, 6—72, 7—120.

Analysis of profiles of these diffraction lines allows us to deduce a structural relationship between the low-temperature phases during the process of ageing. Immediately after quenching, the phase with the tetragonal type of perovskite crystal structure distortions predominates in the bulk of the sample. Then the phase with rhombohedral type of distortions grows to dominate, with some ageing. Only with further ageing the low-temperature phases, namely, the phase with the tetragonal type of lattice distortions and the phase with the rhombohedral type of lattice distortions coexisting as the equilibrium two-phase structure are established in the bulk of the sample.

The "Ti-content–temperature" phase diagrams of the 6/100-Y/Y PLZT and 15/100-Y/Y PLLZT solid solutions are shown in Fig.3.1 and 3.2. As one can see from Fig.3.1 and Fig.3.2 (also Fig.3.8), the phase states of the solid solutions under investigation are two-phase at low ($T < T_C$) temperatures. The domains of the ferroelectric and antiferroelectric phases coexist in the sample volume. It is known that the characteristic time of establishment of the equilibrium low-temperature state (the characteristic time of formation of the order parameter) at the structural phase transitions is of the order of $10^{-4}$–$10^{-6}$ s [54]. The paraelectric–ferroelectric or paraelectric–antiferroelectric phase transitions are also the structural ones. Therefore, all the processes that we investigated in this study (after speed cooling from 600 °C) take place in a system which is essentially a two-phase system containing domains of the coexisting ferroelectric and antiferroelectric phases. Thus, one has to take into account the two-phase nature of the system and the presence of the interphase boundaries while interpreting the results.

The time dependences of the shape and intensity of diffuse X-ray lines (halos), as well as the absence of the said lines in the X-ray patterns obtained at 600°C, confirms the connection between the long-term relaxation and formation of segregates in the vicinity of the interphase boundaries separating the ferroelectric and antiferroelectric phases. The formation of an equilibrium state is a long-term process in solid solutions in which the state of the coexisting domains of the ferroelectric and antiferroelectric phases occurs. As one can see from the X-ray data, it continues for not less than 120 *h*. However, taking into account the limited sensitivity of this method one can assert that this process might take even longer.

The segregation processes are multistage. It is clearly seen from the results presented in Fig. 3.4 – Fig.3.8 that there are different relaxation times caused by different mechanisms. In addition to segregation, one has to note the following circumstance. The processes leading to establishment of equilibrium values of the order parameters during the structural phase transitions occur on the time scale of $10^{-4}$ s (such time intervals are beyond the abilities of our experimental methods). Without elucidation of particular mechanisms responsible for the approach of solid solutions to the equilibrium state, one can assume that the long time constant of this process is connected with the diffusion processes associated with the local decomposition of the solid solutions along the interphase domain boundaries. The estimation of the size of segregates (using the shape of the diffused X-ray lines) gives values of 8–15 *nm* [43, 44] (a similar approach to estimations of the average size of nanoregions one can find in [55-57]).

Long-term relaxation processes are non-monotonic processes due to the condition of 'strong deviation from equilibrium' during initial stages following quenching. In the case of 'weak deviation from equilibrium' (at the final stage) the relaxation process is monotonic and is described by an exponential law. Nonetheless, the PLZT and PLLZT solid solutions do differ by the presence of vacancies in the A-positions of the crystal lattice of the PLZT system of solid solutions. There is a peculiarity of the ageing process for the PLLZT system of solid solutions occurring at ageing times of 20–30 *h*. This peculiarity can be attributed to the accumulation of elastic stress and the subsequent drop in strain [50]. It is not present in the PLZT system. Apart from this the compositional (and structural) relaxation process follows similar patterns for both the PLZT and PLLZT systems of solid solutions.

Now let us discuss mechanisms defining the kinetics of the processes in question at different time periods (stages). There are different mechanisms, which contribute to the long-term relaxation. Two mechanisms should be pointed out among many others. The contribution of the crystal lattice defects, in particular, oxygen vacancies and the diffusion of the cations in the vicinity of the interphase boundaries, caused by local mechanical stresses, command the greatest attention. Under the conditions of our experiments the concentration of vacancies in the lead sites remained practically constant because the volatilization of lead in the PLZT solid solutions starts only at the temperatures above $800°C$. The difference in size (and, consequently, in mobility) and charge of the ions, which are located in the equivalent sites of the crystal lattice, should be taken into account. At the same time a permanent rearrangement of the multiphase domain structure also takes place. This domain structure rearrangement is due to the change of the local composition of solid solution and, as a consequence, due to the change of the local phase stability. Complete analysis is still beyond our grasp because the influence of oxygen sublattice defects on the crystalline structure of these solid solutions is insufficiently characterized at present. However drawing from the experimental results obtained for related oxide materials with the perovskite or perovskite-type structures, some additional insights are possible.

Annealing of samples at $600°C$ promotes growth of the concentration of oxygen vacancies; the equilibrium concentration of oxygen vacancies grows rapidly as the temperature rises. Quenching down to room temperature leads to freezing of the nonequilibrium elevated concentration of vacancies in the bulk. Oxygen vacancies in ionic–covalent compounds, to which the solid solutions with perovskite structure belong, lead to an increase of the crystal lattice parameters [58-61]. In conjunction with the increase in lattice parameter, vacancies in the perovskite and perovskite-like compounds favor the increase of stability of the phases with tetragonal type of crystal lattice distortions, caused by a static $T_2 \times e$ Janh-Teller effect [61, 62] (for example, on $Ti^{3+}$ ions in PZT, $BaTiO_3$ et. al.).

During the ageing process the oxygen vacancies move towards the sample surface and leave the sample (actually, the diffusion of oxygen into the bulk of the sample across the surface takes place) and, as a consequence, the crystal lattice parameters decrease. Since at room temperature the diffusion coefficient is comparatively low, and the surface maintains a steady state concentration of vacancies, the said process is a long-term one.

Let us consider the ion diffusion in the vicinity of the domain boundaries separating domains of the ferroelectric and antiferroelectric phases. Mechanical stresses arise at these interphase boundaries after the quenching and formation of domains of the coexisting ferroelectric and antiferroelectric phases inside the sample's volume. These stresses are caused by the difference in the interplanar distances of neighboring domains. There is an associated increase in the elastic energy. This increase in strain is reduced by the redistribution of ions in the vicinity of the interphase boundaries and, as a result, by the local decomposition of solid solution and formation of segregates. Since mechanical stresses are now the motive force of the ion diffusion, this process must have a higher rate than the process of establishment of equilibrium concentration of the oxygen vacancies. As is seen from Fig. 3.6c, the ageing time dependence of intensity of the diffuse X-ray lines, connected with the formation of segregates, reaches the saturation already in 20–25 $h$. Ions that differ markedly in ionic radii (and charge) participate strongly in the diffusion processes, which define the formation of segregates and the local decomposition. These are lead ions $Pb^{2+}$ and lanthanum ions $La^{3+}$ which occupy the A-sites of the perovskite crystal lattice, and zirconium $Zr^{4+}$ and titanium ions $Ti^{4+}$, which occupy the B-sites of the crystal lattice. Obviously, the rates of their diffusion differ as well. This leads to the change of chemical composition both of the segregates and the solid solutions inside the domains with time. When the ions with smaller ionic radii from the domains of one of the coexisting phases reach the interphase boundary, the solid solution inside these domains becomes enriched with 'larger' ions. As a consequence, the position of the solid solution in the Ti-content–temperature phase diagram changes in time, the crystal lattice parameters increase and the type of the crystal lattice distortion changes. This is clearly seen in Fig. 3.6. When at the first stage of the local decomposition 'small' lanthanum and titanium ions reach the interphase boundaries, the composition of the solid solution inside the domains correspond to PZT with an elevated content of zirconium. Such solid solutions are characterized by the rhombohedral type of crystal lattice distortion. Therefore, as the intensity of the diffuse lines grows, the X-ray diffraction line profiles change. These changes point to the fact that the predominating amount of the tetragonal phase is replaced by that of the rhombohedral phase. That is, the share of phase with the rhombohedral type of distortions increases in the bulk of the sample.

At this stage of the process changes of the elementary cell volume are defined by two competing processes, namely, the reduction of volume owing to the decrease of the concentration of oxygen vacancies and the increase of this volume due to the enrichment of the composition inside domains with the 'larger' ions. The diffusion of oxygen vacancies is the slower process. As a result, during the first stage of ageing the volume increases and a maximum value is achieved in approximately 20 $h$.

As the further ageing takes place, the 'larger' zirconium and lead ions reach the interphase boundaries and the composition inside the domains approaches its nominal formula composition; the solid solution regains its stable location on the temperature versus composition phase diagram (that is, returns to the two-phase region of the state diagram). This process of returning to the composition corresponding to the two-phase region in the phase diagram is clearly manifested in the profiles of the X-ray lines (after approximately 50–60 *h* of ageing) by gradual establishment of precisely such shape which is typical of the domains after sintering and ageing during a very long time (the line shape that is established after one year).

The changes in the profiles of diffusive scattering lines also take place during the above-described ageing process. These changes confirm the fact that the concentration of different ions in the segregates at the interphase boundaries varies. The profile of the diffusive lines is defined by the resulting enveloping curve obtained after summation of the X-ray scattering from the crystal planes in the segregates. Since the local chemical composition of segregates constantly changes in the process of ageing after quenching, the profiles of the diffusive lines changes as well.

The results reported in [47] demonstrate that it is difficult to identify the crystalline structure of the PLZT solid solutions with 6 at.% La, at the temperatures lower than that of the maximum of the $\varepsilon(T)$ dependence in the transient region of the Ti-composition-temperature phase diagram. This is connected with the fact that the domains of the ferroelectric and antiferroelectric phases coexist in the bulk of the sample of those PLZT solid solutions, which belong to the shaded region in the temperature-composition phase diagram, and the effects caused by the interaction of these domains[4, 20, 63] manifest themselves. The coexistence of the ferroelectric and antiferroelectric phases also leads to certain peculiarities in the physical properties of solid solutions located in the above-mentioned region of phase diagram. These peculiarities distinguish the substances with the coexisting ferroelectric and antiferroelectric phases from ordinary ferroelectrics or antiferroelectrics. These peculiarities are manifested, for example, in the dielectric or electro-optical hysteresis loops, in the dispersion of dielectric permittivity in the vicinity of the transition into the paraelectric phase, and in the diffuse nature of this phase transition [5].

As we have indicated, there are many difficulties in the identification of the crystal structure that are connected with the local decomposition of solid solutions in the vicinity of the boundaries between the coexisting phases. The formation of the mesoscopic structure of segregates at these boundaries in the bulk of the samples poses a particular problem. In the X-ray diffraction patterns it manifests itself by the appearance of supplementary diffuse lines that accompany the diffraction lines, based on which the PLZT crystal structure is actually identified. These diffuse lines may often appear as the satellites of the main Bragg peaks. In this case, in the process of mathematical treatment of the experimental X-ray diffraction patterns according to the existing programs a third phase is often brought into consideration. We have faced such a phenomenon while treating experimental results. This is connected only with the fact that in the development of the corresponding computer programs for mathematical treatment of the X-ray data the phenomenon of the local decomposition of solid solution has not been taken into account. As far as we know, the effects associated with the coexistence of ferroelectric and antiferroelectric phase domains and the local decomposition of solid solution in the vicinity of the interphase boundaries have not been discussed in the literature up to now.

In general, the structural analysis of the PLZT solid solutions gets more difficult with the increase of La-content. The increase of La-content reduces the energy barrier that separates the free energy minima corresponding to the ferroelectric and antiferroelectric states. Therefore, the interaction between these phases manifests itself significantly. The boundary (see Fig.1.2) region in the 'Ti-content–temperature' diagram becomes wider, and the degree of crystal lattice distortion decreases, so the splitting of X-ray lines is less pronounced. One more important circumstance is noteworthy. The morphotropic phase boundary originally located at the point approximately corresponding to the composition Zr/Ti = 53/47 [9] in the "Ti-content–temperature" diagram of PZT is now shifted towards solid solutions with higher concentrations of Zr as the La concentration increases. In the PLZT with 8 at.% La the situation is quite complicated. At 8–9% La, the morphotropic boundary is observed in the vicinity of the 65/35 Zr/Ti composition [64]. This composition corresponds to the solid solutions called "relaxor ferroelectrics". Materials with such a composition contain three phases, and their presence is evident from the observed complex structure of the X-ray diffraction lines. Each of the pertinent X-ray lines is a superposition of the expected lines for the ferroelectric rhombohedral, ferroelectric tetragonal and antiferroelectric tetragonal phases. As far as we know, this fact has not been considered during the identification of the crystalline structure of PLZT and, consequently, in the decomposition of X-ray diffraction lines into simple components. Here we would like to mention that transmission electron microscopy investigations [24, 65, 66] of PLZT show that the size of domains of the coexisting ferroelectric and antiferroelectric phases is of the order of 20–30 *nm*. As a consequence an additional broadening of the X-ray diffraction lines is possible.

The identification of the PLZT crystal structure is also affected by the sample preparation method. Such is the case for the hot pressing method used for the preparation of some samples. According to [67], the hot pressing prevents the achievement of a high degree of homogeneity due to such factors as a violation of stoichiometry resulting from hot pressing, an 'underannealing' effects caused by low temperatures applied in the hot pressing process, and a presence of the residual mechanical stresses arising with hot pressing. Even in the case when the hot-pressed PLZT samples possess a high optical quality, some nanometer-scale regions, containing chemical elements that have not reacted completely in the process of PLZT fabrication, are present in the samples' volume. In particular, this fact was confirmed in [68] by transmission electron microscopy. As demonstrated in [69-72] hot-pressed PLZT samples of high optical quality also contain nanodomains with a composition close to that of pure $PbZr_{1-y}Ti_yO_3$.

The results presented here and in [5, 46] explicitly demonstrate that the crystal structure of the PLZT series with 6% La at temperatures below the Curie temperature is complicated. One can identify it with certainty only if solid solutions from different regions of the composition-temperature phase diagram are investigated simultaneously. There is a considerable probability of error in the structure identification when only one solid solution belonging to the boundary region of the composition-temperature phase diagram (dashed region in Fig.3.1) is studied. It is also worth mentioning that the coexistence of ferroelectric and antiferroelectric phases or even more complicated three-phase structure of the PLZT with La-content greater than 6% could be the reason for extraordinary behavior at the phase transition from the paraelectric to the dipole-ordered phase [73, 74].

In conclusion, let us consider the peculiarity of time dependence of the elementary cell volume in the PLLZT solid solutions that manifests itself at aging time intervals of 22 h. The Debye–Scherrer X-ray patterns obtained for the 10/77/23 PLLZT solid solution after different aging time intervals are presented in Fig.3.9. A pronounced complex structure of the (200) X-ray line is clearly seen for the aging time interval of 22 h. We carried out the estimation of sizes of the regions of coherent scattering after different aging time intervals, namely, for $\tau = 22$ h and $\tau = 48$ h for comparison. In the first case, the size of the above regions was $3 \cdot 10^{-4}$ *cm*, and in the second case, it was $1.5 \cdot 10^{-4}$ *cm*. It has to be noted that the regions of coherent scattering are non-uniform and this circumstance leads to the oblong shape of reflexes in diffraction pattern obtained after an aging time of 22 h. The ratio of the long *a* and short *b* axes of the coherent scattering region is $a/b \approx 2.25$. In the case of $\tau = 48$ h the shape of the coherent scattering regions is close to equiaxial.

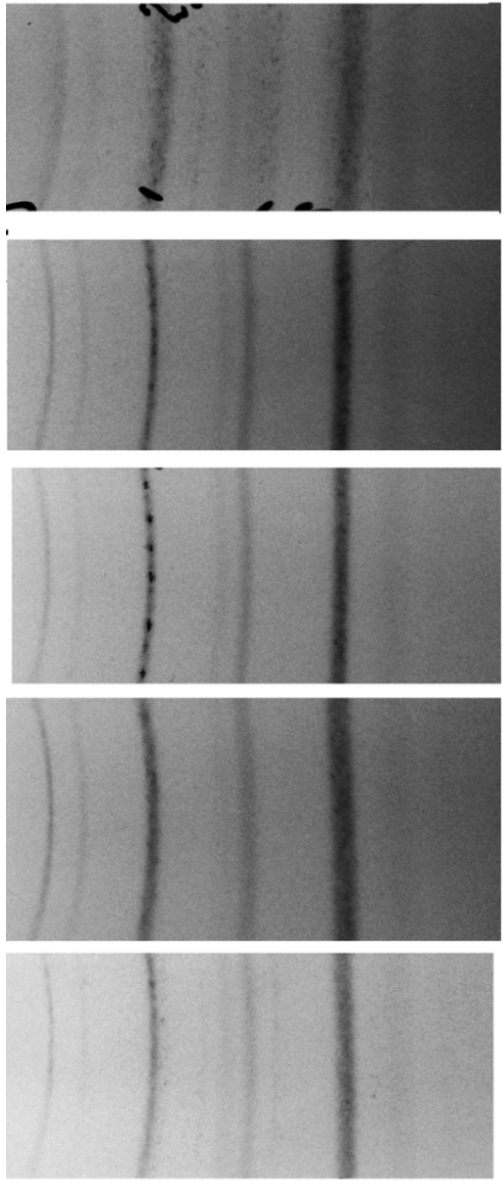

Fig.3.9. X-ray diffraction patterns for the 10/77/23 PLLZT solid solution, obtained after different aging time intervals (hours): a – 0.5, b – 3.0, c – 22.0, d - 48, e - 96.

The complex structure of the (200) X-ray line is the evidence of the presence of texture in the samples after the aging time interval $\tau = 22$ h. This texture was absent at the shorter intervals of aging time. The said texture disappears again at $\tau = 48$ h and longer intervals of aging time. The complex structure of X-ray lines can be attributed to the presence of the internal mechanical stress and to the increase of elastic energy caused by this stress. Such processes are well known and studied for the process of alloy aging (accompanied by the process of local decomposition) [75]. The relaxation toward the equilibrium stress-free state (with smaller value of the elastic energy) takes place by means of ordering in the system of segregates-precipitates including appearance of a texture in the samples.

We attribute the peculiarity in dependence $V(\tau)$ at $\tau = 22$ h to the appearance of internal mechanical stresses, which are a consequence of the considered-above diffusion processes accompanied by the local changes of chemical composition of the solid solution. This peculiarity in the $V(\tau)$ dependence disappears in the process of further aging because of the relaxation of mechanical stresses. The absence of such peculiarity in the $V(\tau)$ dependence for the PLZT solid solution is explained by the fact that mechanical stresses relax by means of redistribution of vacancies in the regions where mechanical stress appears. Let us remind that in the PLZT solid solutions (also investigated in this paper) each two lanthanum ions lead to appearance of one additional vacancy in A-cites of the crystal lattice.

Thus, in spite of the similarity of the processes of local decomposition in the PLZT and PLLZT solid solutions some differences are connected with the essentially different number of vacancies in the crystal lattice in these systems of solid solutions and therefore with the presence of additional mechanical stresses in the PLLZT solid solutions.

## 4. Some effects caused by the interfaces between domains of the coexisting ferroelectric and antiferroelectric phases

**4.1.** *Ferroelectric-antiferroelectric interfaces and formation of textured structures*

In the substance with the system of coexisting domains of two dipole-ordered phases there exists the system of interphase boundaries between these domains. By virtue of this system of interphase boundaries in the vicinity of which the local decomposition of the solid solution takes place one can fabricate various structures (the textured ones among them). These structures lead to new uncharacteristic properties of habitual materials. Due to the fact that the decomposition of the solid solution and the segregation in the vicinity of the interphase boundaries are long-term processes they can be controlled by means of different external influences. The easiest way to accomplish this control is at the temperatures corresponding to the paraelectric state.

As it is shown during theoretical consideration of the phase coexistence in Chapter 2, the formation of the two-phase (ferroelectric + antiferroelectric) domains takes place in the paraelectric phase at the temperatures substantially higher than the Curie temperature. It has been proved in experiments [23] in which the dependences of the shapes of X-ray diffraction lines on temperature at $T > T_c$ (in the paraelectric phase) for the 10/100-Y/Y PLLZT solid solutions have been investigated. The (222) and (400) X-ray lines, the most typical for perovskite crystal structure, have been chosen for this purpose. The former line is singlet in the case of the tetragonal crystal lattice distortions and in the case of the rhombohedral type of lattice a distortions this line is doublet. The latter line is a doublet when the distortions are of the tetragonal type, and this line is a singlet when the distortions are rhombohedral. The analysis of the shapes of these X-ray diffraction lines allowed to demonstrate which of the low-temperature phases were extended to the high-temperature region, situated above the point of the paraelectric phase transition for the main part of the samples' volume.

The temperature dependences of the asymmetry of the X-ray lines ($\gamma = S_l/S_r$) and the crystal cell parameters for the PLLZT solid solutions belonging to different regions of the "Ti-content-temperature" phase diagram are given in Fig.4.1. Here $S_l$ and $S_r$ are the areas below the contour of the X-ray diffraction line located from the left and from the right of the vertical line drawn from the vertex of the X-ray line, $S = S_l + S_r$, where $S$ is the total area of the X-ray diffraction line.

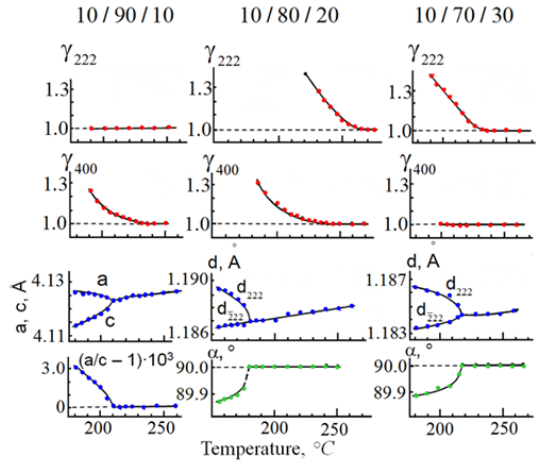

Fig.4.1. Temperature dependences of X-ray parameters for 10/90/10 (on the left), 10/80/20 (in the middle) and 10/70/30 (on the right) PLLZT solid solutions [23, 73].

According to the calorimetric, dielectric, and X-ray diffraction data, the 10/90/10 PLLZT undergoes paraelectric phase transition at 209°C. The temperature dependences of the crystal lattice parameters $a$, $c$ and ($a/c - 1$) near the point of phase transition are shown in the left column in Fig.4.1. The asymmetry of the (400) line is preserved up to temperatures of 230°C. The (222) X-ray diffraction line is symmetric for temperatures above 209°C. This fact directly testifies that the diffuseness of the phase transition in this solid solution is connected with the extension of domains of the tetragonally distorted antiferroelectric phase into the range of high-temperatures above the transition point.

The 10/70/30 PLLZT solid solution undergoes the paraelectric phase transition at 218°C. The temperature dependences of the crystal lattice parameters near the point of phase transition are shown in the right column in Fig.4.1. The asymmetry of the (222) line is preserved at temperatures up to 235°C. The (400) X-ray diffraction line is symmetric within the entire interval of temperatures above 218°C. This fact shows that the diffuseness of the phase transition in this solid solution is connected with the extension of the domains of rhombohedrally distorted ferroelectric phase into the high-temperature region above the transition point.

For the 10/80/20 PLLZT solid solution, located closely to the boundary between the regions of the ferroelectric and antiferroelectric ordering in the "Ti-content-temperature" phase diagram the observed picture differs from those described above. In this case both the (222) and (400) X-ray diffraction lines are asymmetric in the temperature region above the transition point (these data are shown in the middle column in Fig.4.1). The temperature interval, within which the asymmetry of X-ray lines is preserved, is much wider than those temperature intervals for the 10/90/10 and 10/70/30 PLLZT. The main part of the samples' volume undergoes phase transition at 180°C, whereas the complete symmetry of the X-ray diffraction patterns becomes obvious only at temperatures above 260°C.

A comprehensive analysis of the profiles of the X-ray diffraction lines, at the temperatures above the Curie point revealed that the samples have a complex phase composition. In the first approximation, it may be concluded that the paraelectric matrix of the sample as a whole contains the two-phase domains which, in their turn, consist of the adjoining domains with the ferroelectric and antiferroelectric ordering.

In [74] a creation of two-phase domains was realized using a DC electric field. Electric field was applied to samples during 1 h at the temperatures that exceeded $T_c$ by more than 50 – 100 °C. After that the field was switch out, and the cooling of samples to room temperature was carried out without electric field. Fig.4.2 shows dependences of the piezoelectric coefficient $d_{33}$ on the Ti-content after described procedure for 10/100–Y/Y PLLZT solid solutions.

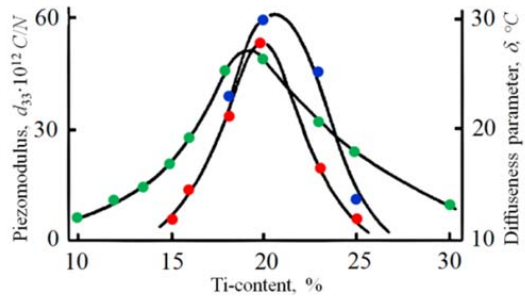

Fig.4.2. Dependences of piezoelectric modules (red and blue dots) on titanium content for the 10/100-Y/Y PLLZT series of solid solutions measured after different regimes of thermal treatment in paraelectric phase. Dependence of diffuseness parameter $\delta$ (green dots) on titanium content for the 10/100-Y/Y PLLZT series of solid solutions

As it is easily seen, the maximum values of the piezoelectric coefficient are observed for solid solutions with the compositions 10/80/20 and 10/77/23. These solid solutions are characterized by the coexistence of the ferroelectric and antiferroelectric phases in the sample's volume. The value of $d_{33}$ coefficient decreases rapidly when the composition of the solid solution is shifted toward the regions of both pure ferroelectric and antiferroelectric states. It is zero for both 10/90/10 and 10/70/30 solid solutions. After cooling of the samples without DC electric field piezoelectric coefficient $d_{33}$ is zero always. In order to demonstrate that the diffuseness of the paraelectric phase transition is also dependent on the presence of the coexisting ferroelectric and antiferroelectric phases we included in Fig. 4.2 the dependence of the diffuseness parameter, $\delta$, (the definition of the diffuseness parameter on can find in [76]) on the Ti-content in the PLLZT solid solitions.

**4.2.** *Dielectric memory*

Phenomena of the so-called thermal dielectric, field, or mechanical memory that had been experimentally observed in the PLZT solid solutions are of interest from the physical point of view. All above mentioned effects have common physical basis, namely, the local decomposition of solid solution in the vicinity of interdomain walls. Therefore, here the phenomenon of thermal dielectric memory will only be discussed in detail. This effect implies that when after annealing at high temperatures the samples are cooled to some particular temperature $T = T_{age}$ and are subjected to aging at this temperature (during the time interval of 20 hours and longer) a specific characteristic feature is observed on the $\varepsilon(T)$ dependencies at the cycling after aging. This feature consist in a "drop-shaped" behavior of the dependence $\varepsilon(T)$ in the vicinity of the aging temperature $T_{age}$. Typical results of experiments are presented in Fig.4.3 [77] and Fig.4.4 [78].

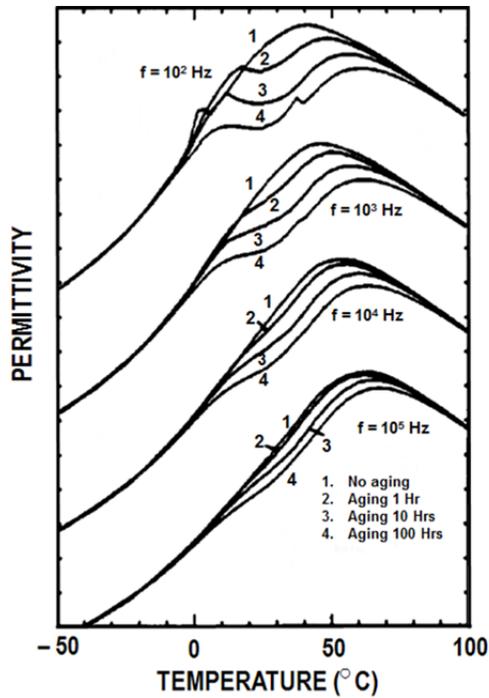

Fig.4.3. Temperature dependences of dielectric constant for different aging time at ~ 23 °C for the 9.5/65/35 PLZT annealed at 400 °C [77].

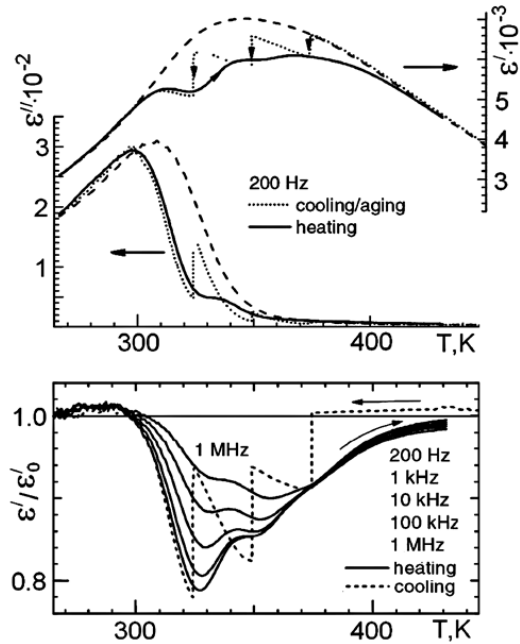

Fig.4.4. Dielectric permittivity of the 9/65/35 PLZT at 200 Hz during multiple aging stages (24 h each) and subsequent heating curves with memory, compared with the reference curve measured on continuous cooling [78].

All presented results have simple unambiguous explanation if one takes into account the inhomogeneous structure of coexisting domains of the ferroelectric and antiferroelectric phases existing in the sample. Domain structure is formed in the bulk of the sample after the high-temperature annealing. The shares of each phase (and consequently the domain sizes) are determined by the relation between the free energies of these phases. Diffusive local decomposition of the solid solution constantly takes place in the vicinity of the interdomain boundaries. However, this is the long-time process and it is practically not manifested in material parameters. This influence is extremely small due to the fact that the characteristic rate of change of temperature (during the temperature measurements) is of the order of several degrees Celsius per minute. Thus, the characteristic time of temperature measurements is much shrter than the characteristic relaxation times of diffusion processes (which are of the order of tens of hours). The same is true for the external field or mechanical stress measurements. During aging at the temperature $T_{age}$, which lasts for long time (tens of hours), the local decomposition of the solid solution takes place and the structure of the segregates formed in the sample repeats the structure of the interphase boundaries. The longer is the aging time the more pronounced is the spatial structure (the distinctive network) of the segregates and the larger is the volume of the sample occupied by these segregates. The dielectric permittivity is reduced as a result (Fig.4.3 and Fig.4.4). The spatial structure of segregates formed during the sample's aging is conserved during the temperature cycling (with the rate of several degrees per minute) following the aging.

The equilibrium relation between the shares of each phase (ferroelectric and antiferroelectric) and the sizes of domains of these phases are constantly changing in the process of temperature cycling. The domain sizes and the structure of domains coincide with the preserved spatial structure of segregates when the temperature $T_{age}$ is achieved. At this moment the effective pining of the interphase boundaries takes place and the contribution of the oscillations of interdomain boundaries determining the main contribution to the dielectric permittivity decreases. This pining of the interdomain boundaries leads to the appearance of the "drop-shaped" feature on the temperature dependence of dielectric constant, $\varepsilon(T)$.

The nature of the effect of dielectric memory at the temperatures both above $T_m$ and below $T_m$ is the same as one can clearly see from the results of [78] (see also [79]). However, the authors of numerous papers who were trying to connect this phenomenon with relaxor ferroelectrics were forced to come up with different mechanisms for these two cases. One does not need to do that using the model suggested here. As shown during the model consideration (see the part 2), the structure of coexisting domains of the ferroelectric and antiferroelectric phases is present both above $T_m$ and below $T_m$ (let us remind that the X-ray studies carried out above $T_m$ demonstrated the presence of the two-phase (ferroelectric + antiferroelectric) domains formed by the coexisting ferroelectric and antiferroelectric phases [23, 73, 74]). Thus, the interdomain wall structure that determines the dielectric memory effect is present both below $T_m$ and above $T_m$.

The shares of the antiferroelectric and ferroelectric phases and, thus, the spatial structure of interdomain walls can be also changed by means of other thermodynamic parameters such as an external field and a mechanical stress. The long time aging of the samples at some nonzero values of the above-mentioned parameters will lead to formation of the new spatial structure of segregates and the effects of the electric filed memory or the mechanical dielectric memory will be manifested during the cyclic changes of these parameters after the aging [80-82]. These effects are presented in Fig.4.5 and Fig.4.6.

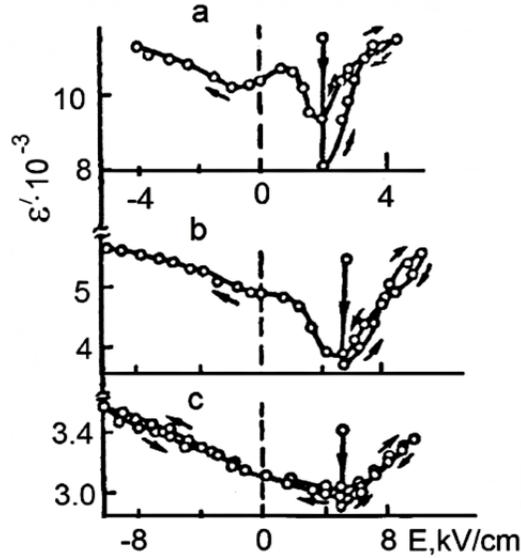

Fig.4.5. Dielectric constant vs. DC electric field in the PLZT solid solutions after aging during 20 h [80]. The solid solutions compositions: a – 8/65/35, $T_{age}$ = 57 °C; b – 11/65/35, $T_{age}$ = 22 °C; c – 13.5/65/35, $T_{age}$ = 22 °C.

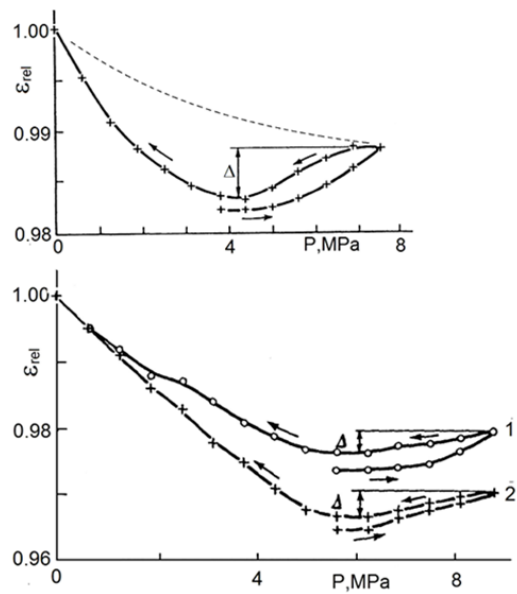

Fig.4.6. Dependences of the normalized dielectric constant on uniaxial stress for the 8/65/35 PLZT after aging during 20 h [81] (at the top). The dashed line corresponds to the ε(P) dependence, obtained without aging (added by authors of the present paper). Dependences of the normalized dielectric constant on uniaxial stress after aging during 48 h (1) and 120 h (2) in a DC electric field $E$ = 1850 $V/cm$, ($T_{age}$ = 27 °C, $P_{age}$ = 3.8 MPa) for the 8/65/35 PLZT solid solution (on the bottom).

The effects discussed above are as a rule attributed to the so-called dipole-glass-like behavior in relaxor systems. However, based on this dipole-glass approach it is impossible to describe and explain all manifestation of a number of other phenomena observed in relaxor ferroelectrics. We demonstrated that rather simple explanation can be given if one takes into account coexistence of the antiferroelectric and ferroelectric phases.

**4.3.** *Peculiarities of dielectric properties. Dielectric relaxation*

It was shown in [63, 84] that in systems with close values of the free energies of the ferroelectric and antiferroelectric phases and the coexisting domains of these phases the action of the DC electric field on such systems leads to the change of the state of the substance due to the displacement of interdomain walls while the internal state within domains is being unchanged. This motion of interdomain walls is inertial and is accompanied by relaxation processes. The relaxational dynamics of the interdomain walls should manifest itself most vividly when the said substances are subjected to the action of an AC electric field.

Main features of the interdomain wall dynamics under action of an AC electric field can be obtained by examining the forces acting on the interdomain wall. The basic equation derived from the condition of balance of forces affecting the interdomain wall, has the form:

$$P_{E(t)} + R(u) + P(u,t) = 0, \qquad (4.1)$$

where $u = u(t)$ is the time-dependent displacement of the interdomain wall, $E(t)$ is the electric field intensity; $P_{E(t)}$ is the pressure acting on the interdomain wall associated with the expansion of the volume of ferroelectric domain in external field. The forces which counteract the displacement of the interdomain wall under the effect of the field can be presented as a sum of two terms: $R(u)$ is the pressure of the forces caused by the interaction of the interdomain wall with immobile defects of the crystal structure, and $P(u,t)$ is the pressure of the so-called after-effect forces. The latter forces may be caused by different factors: the interaction of the interdomain wall with mobile defects of the crystal lattice, as well as nonzero duration of the thermal processes connected with the phase transitions in those regions of the crystal through which the interdomain wall passes.

Eq. (4.1) describes the equilibrium of the interdomain wall under the action of different kinds of physical forces. Considering this relation as a motion equation (Newton equation), one can easily see that here the terms $m_w \ddot{u}(t)$ ($m_w$ is the effective mass of the interdomain wall) and $\gamma m_w \dot{u}(t)$ are not taken into account. These terms may be neglected during the analysis of motion when the effective mass $m_w$ is small and when the frequency of an AC electric field is essentially lower than the frequency of the interdomain wall oscillations in the potential well $\int du R(u)$ created by immobile crystal lattice defects.

The effective pressure exerted on interdomain wall by an external electric field can be presented in the form:

$$P_{E(t)} = E(t) P_s. \qquad (4.2)$$

For simplicity we shall consider a uniaxial ferroelectric, and the direction of external field coinciding with the polar axis.

The pressure of the force affecting interdomain wall from static defects can be presented by the first term of the power series expansion in displacements (i.e. by a quasi-elastic force):

$$R(u) = -\alpha u, \quad (\alpha > 0). \qquad (4.3)$$

The pressure $P(u,t)$ is defined by the entire prehistory of the sample under investigation. Therefore it must depend on the values of $u$ at any time $t'$ preceding the present measurement time $t$ ($t' < t$).

Analytically this dependence can be chosen at different degree of complexity. Here we choose it in the simplest form. From the most general consideration, $P(u,t)$ can be presented as

$$P(u,t) = -w\int_0^t F[u(t),u(t')]g(t-t')dt', \qquad (4.4)$$

where

$$F[u(t),u(t')] = F[u(t)-u(t')],$$

$g(t-t')$ is a function of aftereffect which describes the contribution of the relaxation effects existing at the time $t'$ ($0 < t' < t$) to the resultant relaxation at the moment of time $t$. For a system with one relaxation time the function $g(t-t')$ has a simple form:

$$g(t-t') = \frac{1}{\tau}\exp\left(\frac{t-t'}{\tau}\right). \qquad (4.5)$$

Further, by analogy with (4.3), for small displacements of the interdomain wall the function $F[u(t) - u(t')]$ can be presented in a linearized form:

$$F[u(t)-u(t')] = -k[u(t)-u(t')]. \qquad (4.6)$$

Thus, the expression describing the pressure of aftereffect acquires the form:

$$P(u,t) = kw\int_0^t [u(t)-u(t')]\cdot\exp[-(t-t')/\tau]\frac{dt'}{\tau}. \qquad (4.7)$$

Taking into account the above-mentioned approximations one can present the equation for the interdomain wall motion in the following form:

$$-\alpha u(t) + kw\int_0^t \frac{dt'}{\tau}[u(t)-u(t')]\exp\left(-\frac{(t-t')}{\tau}\right) + P_s E_0 \exp(i\omega t) = 0 \qquad (4.8)$$

or

$$[1+\eta G(t)]u(t) - \eta\exp\left(-\frac{t}{\tau}\right)\int_0^t u(t')\exp\left(\frac{t'}{\tau}\right)\frac{dt'}{\tau} = u_0 \exp(i\omega t),$$

where $\eta = kw/\alpha$, $u_0 = P_s E_0/\alpha$ is the interdomain wall displacement under the action of a DC electric field $E_0$ and $G(t) = (1-\exp(-t/\tau))$ is the time function. This equation should be solved either by a method of successive approximations (in the first approximation) or by transforming it into a differential equation which can be easily solved in the stationary case $t \to \infty$ ($G(t) = 1$). Representation of the steady-state solution in the form $u(t) = \tilde{u}\exp(i\omega t)$ leads to the following expression for the amplitude $\tilde{u}$ [85]:

$$\tilde{u} = u_0 \frac{(1+i\omega\tau)}{[1+i\omega\tau(1+G(t)\eta)]} \to u_0 \frac{(1+i\omega\tau)}{[1+i\omega\tau(1+\eta)]}. \tag{4.9}$$

The interdomain wall displacements under the action of the electric field lead to a change of the dipole moment in the volume of the substance involved by this displacement. It gives an easy way to determine the dielectric susceptibility $\chi$ connected with this process:

$$\chi(t) = P_s S u(t) / E(t), \tag{4.10}$$

where $S$ is the area of interdomain wall. Let $\chi_0 = P_s^2 S / \alpha$ denotes the static susceptibility then using equations (4.10) and (4.9) one can obtain:

$$\chi = \chi' - i\chi'' = \chi_0 \frac{1+\omega^2\tau^2(1+\eta) - i\omega\tau\eta}{1+\omega^2\tau^2(1+\eta)^2}. \tag{4.11}$$

From here it follows that:

$$\chi'(\omega, T) = \chi_0 \frac{1+\omega^2\tau^2(1+\eta)}{1+\omega^2\tau^2(1+\eta)^2} \equiv \chi_0(T)F_1(\omega, T)$$

$$\chi''(\omega, T) = \chi_0 \frac{\omega\tau\eta}{1+\omega^2\tau^2(1+\eta)^2} \equiv \chi_0(T)F_2(\omega, T) \tag{4.12}$$

As noted earlier, the relaxational dynamics of interdomain walls is influenced by the aftereffect of the first-order phase transition between the ferroelectric and antiferroelectric states, happening in that part of the sample where the interdomain walls motion takes place. In this case the said states are separated by a potential barrier, and the thermoactivation processes take place. Therefore it is natural to assume that the temperature dependence of the relaxation time has the form:

$$\tau = \tau_0 \exp(\Delta / kT). \tag{4.13}$$

Here $\Delta$ is the activation energy, $\tau_0$ is the reciprocal value of the thermal activation frequency (its order of magnitude equals $10^{-11} - 10^{-13}$ s).

As shown in [63], the coexistence of domains of the ferroelectric and antiferroelectric phases is provided by the equality of their thermodynamic potentials within a wide interval of the electric field intensity. The change in the intensity of the external field results in the displacement of the interphase boundary, the energy of each of the coexisting phases remains unchanged ($\Delta = 0$). However, the presence of the effect of the "interdomain wall lag", caused by the aftereffect forces (4.4), leads to changes in this simple picture. Now the condition of the field compensation inside the domains is not fulfilled, and $\Delta \neq 0$ (though being a small value). Under the condition $\omega\tau \ll 1$ the functions $F_1(\omega,T)$ and $F_2(\omega,T)$ acquire the following forms:

$$F_1(\omega,T) \approx 1 - \omega^2\tau^2\eta(1+\eta) \approx 1 - 2\omega^2\tau_0^2\eta(1+\eta)\Delta/kT \tag{4.14}$$

$$F_2(\omega,T) \approx \eta\omega\tau \approx \eta\omega\tau_0(1+\Delta/kT)$$

It was shown [86, 87], that under the condition of $\omega\tau \ll 1$, expressions (4.12) describe the shift of $T'_m$ and $T''_m$, caused by the change of the field frequency ($T'_m$ is the temperature of the maximum of $\varepsilon'(T)$ dependence and $T''_m$ is the temperature of the maximum of $\varepsilon''(T)$ dependence). In particular, substituting (4.14) into expressions (4.12) and calculating a derivative with respect to $(1/T)$ and equating it to zero, one can obtain the following expression for $T'_m$:

$$\left(\frac{1}{T'_m}\right) = -\frac{k}{\Delta}\left[\ln\omega + \ln\tau_0 - V\left(\frac{1}{T'_m}\right)\right] \tag{4.15}$$

where $V(1/T)$ is a slow varying function. The introduction of the effective temperature $T_f$, i.e. the linearization of (4.15) using the method of Pade approximation [88], yields another form of this expression:

$$\frac{1}{T'_m - T_f} = -\frac{k}{\Delta}[\ln\omega + \ln\tau_0], \tag{4.16}$$

which can be also written down in the form of the well-known Vogel-Fulcher relation:

$$\omega = \left(\frac{1}{\tau_0}\right)\exp\left[-\frac{\Delta}{k(T'_m - T_f)}\right]. \tag{4.16a}$$

The expressions for $T''_m(\omega)$ are analogous to (4.16) and (4.16a).

A mathematical expression which relates the location of the maximum of the $\varepsilon'(T)$ dependence with the electric field amplitude can be easily derived from (4.16). At first, we consider a DC electric field. When the electric field is applied to the sample, eq. (4.13) acquires the following form:

$$\tau = \tau_0 \exp\left(\frac{\Delta}{kT}\right) \xrightarrow{E} \tau_0 \exp\left[\frac{(\Delta - \alpha P_s E)}{kT}\right], \tag{4.17}$$

where the second term in the exponent describes the change of the ferroelectric minimum of thermodynamic potential, caused by the field. The substitution $\Delta \to \Delta - AE$ needs to be made in expression (4.16) to obtain the dependence $T'_m(E)$. It is easily seen that the increase of the field intensity leads to a near linear decrease of $T'_m(E)$.

$$\frac{1}{T'_m - T_f} = -\frac{k}{\Delta - AE}\ln\omega - \frac{k}{\Delta - AE}\ln\tau_0. \tag{4.18}$$

Expression (4.18) can be used for consideration of the AC field influence on the location of $T'_m$. The applicability of (4.17) for the description of the interdomain wall motion in an AC electric field in the scope of this model is ensured by the fact that both the process of relaxation of the order parameters to the steady state under the influence of the field and the change of the height of the potential barrier can be considered simultaneously.

The relaxational dynamics of interdomain walls is determined by the term $P(u,t)$ in (4.1). Let us remind that this term may be caused by different factors: the interaction of the interdomain walls with mobile crystal lattice defects, as well as nonzero duration of the thermal processes connected with phase transitions in those regions of the crystal through which interdomain wall passes. Characteristic time of the above-mentioned processes is rather long. At the same time the processes of establishment of the order parameter during the structural phase transitions (to which the ferroelectric and antiferroelectric phase transitions belong) are the fast ones. Characteristic times of these processes are several orders of magnitude shorter. Based on the assumptions used to write eq. (4.1), one can consider the establishment of the value of the potential barrier (parameter $\Delta$) between two minima corresponding to the ferroelectric and antiferroelectric phases in eq. (4.13) as instantaneous when the AC electric field is applied to the sample (that is the value $\Delta$ follows the instantaneous value of the field). That is why the dependence $T'_m$ on the AC field amplitude will be the same as the dependence on the intensity of the DC field, namely, an increase of the field amplitude will lead to a near linear decrease of $T'_m$.

Now let us discuss experimental results on dielectric properties of the substances with a small difference in the energies of the ferroelectric and antiferroelectric states. As an example, we consider the 10/100-Y/Y PLLZT and the 6/100-Y/Y PLZT series of solid solutions. Phase diagrams of these series of solid solutions are similar to the ones shown in Fig.3.1 where the regions of the coexisting ferroelectric and antiferroelectric phases are clearly seen.

First of all we consider the experiments carried out on the PLLZT solid solutions [83]. The dielectric response obtained using AC field at 1 kHz, with the amplitude varying between 1 and 150 *V/mm*, for the PLLZT solid solutions located in the three different regions of the "Ti-content-temperature" phase diagram is shown in Fig.4.7. It should be noted that for the above-mentioned values of the electric field intensity the dependences $P(E)$ are linear. The effects bound up either with the reorientation of domains of the ferroelectric phase or with the ferroelectric phase induction manifest themselves in fields which noticeably exceed the used ones. As it can be seen, relaxation appears only for solid solutions with coexisting ferroelectric and antiferroelectric phases.

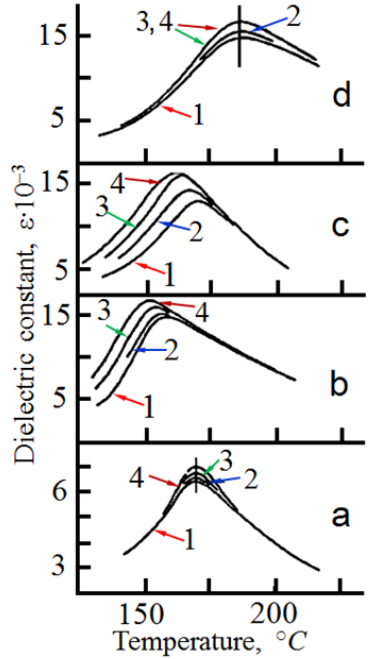

Fig.4.7. Temperature dependences of the real part of dielectric constant for the 10/100-Y/Y PLLZT solid solutions [83, 86].
Ti-content Y, %: a – 15; b – 20; c – 23; d – 30; $E_-$, V/mm: 1 – 1; 2 – 50; 3 – 100; 4 – 150.

The dispersion of dielectric constant was investigated within a frequency range from $10^2$ to $10^5$ Hz. Two series 10/90/10 and 10/70/30 of the PLLZT solid solutions located within the region of the uniform antiferroelectric and ferroelectric states in the "Ti-content-temperature" phase diagram were used for measurements. The real and imaginary parts of the dielectric constant $\varepsilon'(T)$ and $\varepsilon''(T)$ (as well as $T'_m$ and $T''_m$ respectively,) did not manifest dependence on frequency. However, for the 10/80/20 and 10/77/23 PLLZT the changes in the behavior of $\varepsilon'(T)$ turned out to be fundamental (Fig.4.8). The $\varepsilon'(T)$ dependences were displaced towards higher temperatures and $\varepsilon'_m$ noticeably decreased with increasing frequency of the electric field.

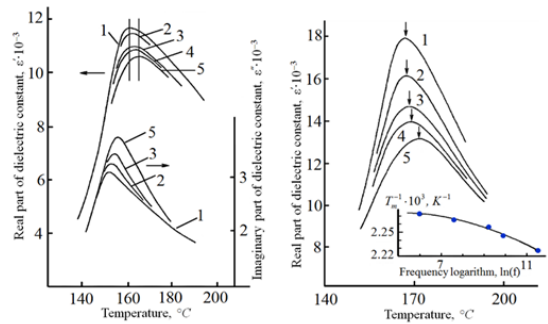

Fig.4.8. Temperature dependences of $\varepsilon'$ and $\varepsilon''$ for the 10/80/20 PLLZT (left) and the temperature dependence of $\varepsilon'$ for the 10/77/23 PLLZT (right) [35, 83, 86]. Frequency (kHz): 1 – 0.4; 2 – 2.0; 3 – 10.0; 4 – 20.0; 5 – 100.0.

The dependences of the imaginary part of dielectric constant, $\varepsilon''(T)$, also changed significantly: $\varepsilon''(T)$ curves were shifted towards higher temperatures and the value of $\varepsilon''_m$ increased. For example, for the 10/77/23 PLLZT the Vogel-Fulcher relation is fulfilled in the following form: $\omega = (1/\tau_0)\exp\left[-\Delta/k(T_m - T_f)\right]$, with the effective temperature $T_f$ = (425 ± 2) K. More detailed discussion of these results one can find in [83, 86].

It is also interesting to mention the studies of effect of a DC bias applied to the samples of materials with coexisting ferroelectric and antiferroelectric phases. These studies were carried out on the PLLZT solid solution with 15% of Ti. The "Electric field-temperature" phase diagram for 10/85/15 PLLZT is presented in Fig.4.9.

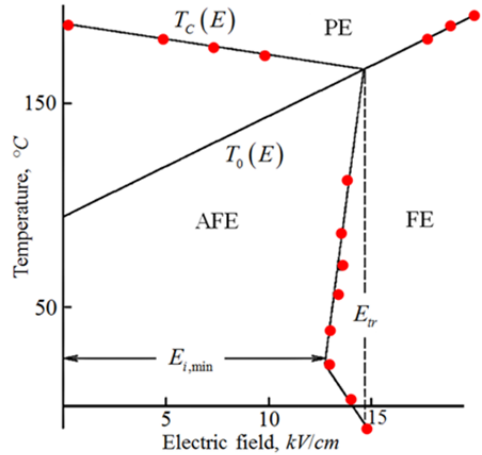

Fig.4.9. "Electric field-temperature" phase diagram for the 10/85/15 PLLZT solid solution [35, 83].

For a low-intensity DC bias only one phase transition from an antiferroelectric phase to a paraelectric phase takes place on the line $T_c(E)$. For a DC bias $E > E_{tr} \cong 15$ kV/cm, the ferroelectric to paraelectric phase transition is observed.

The temperature dependences of the dielectric constant $\varepsilon'(T)$ for different values of DC bias and intensities of the measuring AC fields are presented in Fig.4.10.

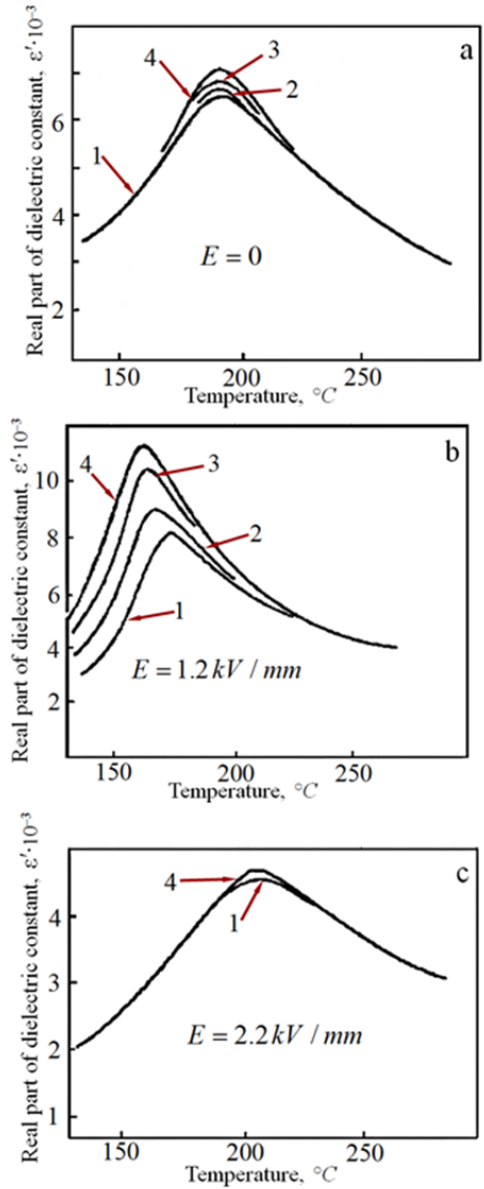

Fig.4.10. Temperature dependences of the real part of dielectric constant, $\varepsilon'$, for the 10/85/15 PLLZT solid solution at different intensities of the AC and DC fields [35, 83]. AC filed amplitude (V/mm): 1 – 1; 2 – 50; 3 – 100; 4 – 150

For zero DC bias (Fig.4.10 (a)) the location of the maximum of $\varepsilon'(T)$ is independent on the AC field amplitude, and the value of $\varepsilon'_m$ somewhat increases with the rise of the AC field. Similar behavior is observed for a DC bias intensity $E$ = 22 kV/cm (Fig.4.10 (c)). The $\varepsilon'(T)$ dependences obtained for different AC field amplitudes are practically coincident in the whole of the temperature range studied.

Different dielectric behavior is observed for the DC bias $E$ = 12 kV/cm (Fig.4.10 (b)). The maximum value of dielectric constant $\varepsilon'_m$ increases and the temperature $T'_m$ decreases when the AC field amplitude rises. Simultaneously, the large increase of the dielectric constant takes place for $T < T_m$.

As seen from Fig.4.7, 4.8 and 4.10, peculiarities in the behavior of permittivity caused by an increase of the amplitude of a measuring AC field are revealed only in the solid solutions located near the boundary separating the regions of the ferroelectric and antiferroelectric states in the "Ti-content-temperature" or the "Electric field-temperature" phase diagrams. These solid solutions have close values of the free energies of the ferroelectric and antiferroelectric states, and this leads to the coexistence of domains with the ferroelectric and antiferroelectric orderings in the volume of the sample. In this case one should take into consideration the contribution of interphase boundary oscillations. As seen from the present results, the deviations of the permittivity from the classic behavior take place only in the case when the contribution of the said boundaries has the form predicted by the model for the ferroelectric-antiferroelectric phase transformations considered above. At the same time, it is not essential which of the external factors – the change of the DC bias intensity or the change of the solid solution composition – results in the appearance of the equilibrium two-phase state.

The difference in behavior of the dielectric constant under changing intensity of the measuring AC field should be especially considered for the case of weak and strong fields (with intensities lower and higher than 20 $V/mm$, respectively). In our opinion, the pinning of the interdomain walls on defects of the crystal structure manifests itself in weak fields. In more detail this problem is discussed in [83].

The relaxation properties of the PLZT solid solutions are analogous to the ones for PLLZT solid solution.

**4.4.** *Tuning piezoelectric properties by means of the phase transition via intermediate state*

Physical characteristics of the substance manifest significant changes during phase transformations. This circumstance may be used to create materials with controlled operational characteristics. Such control can be brought into play by the variation of external thermodynamic parameters that give rise to the phase transformation. However, the main difficulty in using phase transitions is that, as a rule, they occur quickly and within a narrow range of variation of external parameters.

Phase transformations via intermediate state in magnetic materials and superconductors belong to few exceptions to this rule [1, 2]. In some of these substances phase transitions take place in a wide interval of magnetic field. The phase transitions which involve the intermediate state have been studied in detail both theoretically and experimentally in these materials long time ago. In ferroelectrics and antiferroelectrics such transitions have not been revealed for quite a long time, despite the generality of the phenomenological description of phase transitions in all three classes of substances. As shown in [63], the intermediate state analogous to that in magnetic crystals or superconductors cannot exist in ferroelectrics or antiferroelectrics. At the same time, it was demonstrated that, in the case when a ferroelectric state is induced in an antiferroelectric state by the electric field (or an antiferroelectric state is induced in a ferroelectric state by the hydrostatic pressure), an inhomogeneous state of domains of the coexisting ferroelectrics and antiferroelectrics phases may appear. Such state is similar to the intermediate state in magnetic materials or superconductors as far as its geometrical aspects are concerned.

To consider the system of coexisting ferroelectric and antiferroelectric phases subjected to the action of a DC electric field, one has to make a changeover $E_{\eta_{\alpha',i}} \to E_{\eta_{\alpha',i}} - E_{\alpha,i;ext} \equiv E_{\alpha,i;\text{int}}$ (here $E_{\alpha,i;ext}$ and $E_{\alpha,i;\text{int}}$ are external and internal electric fields, respectively) in (2.2), (2.5), and (2.6). Equations (2.5) and (2.6) now acquire the form:

$$\xi_\alpha \left( \frac{\partial \varphi_\alpha}{\partial \eta_{\alpha,i}} + E_{\eta_{\alpha',i}} - E_{\alpha,i;ext} \right) = 0, \quad (\xi_\alpha \neq 0) \tag{4.19}$$

$$\varphi_\alpha + \eta_{\alpha,i} E_{\alpha,i;\text{int}} = \lambda = const,  \qquad (4.20)$$

As seen from eq. (4.20), the condition for coexistence of the thermodynamically equilibrium multiphase structure is the equality of thermodynamic potentials of the phases, taking into account the external and internal effective fields. The intrinsic fields $E_{\eta_{\alpha,i}}$ are spatially varied. Therefore, at the same value of the external field $E_{\alpha,i;ext}$ the phase transition may take place only in certain local regions of the sample (but not in the whole of the sample). This means that, within a certain interval of values of the external electric field the domains of the phases, participating in the transition, coexist in the bulk of the sample. By analogy with magnets or superconductors, such state is called the intermediate state in antiferroelectrics.

In the case of the antiferroelectric to ferroelectric phase transition, the share of the ferroelectric phase increases linearly within the range of values of the electric field where the intermediate state takes place [46]:

$$\xi_1 \cong \frac{E_{ext} - E_{pt}}{C_1 P_{1,0}(E)}. \qquad (4.21)$$

The boundaries of the intermediate state (the electric field intensities for the onset and the end of the transition) are defined by the equations:

$$E_1 \cong E_{pt}; \qquad E_2 \cong E_{pt} + C_1 P_{1,0}(E_{pt}). \qquad (4.22)$$

In two last equations $E_{pt}$ is the electric field intensity at which the phase transition from the antiferroelectric to ferroelectric state takes place when the interphase interaction is absent, $P_{1,0}(E_{pt})$ is the polarization of the ferroelectric phase when the intensity of an applied DC field equals to $E_{pt}$ ($P_{1,0}(E) \approx P_0$) and $C_1$ is the coefficient that appears in equations (2.3) and (2.9).

The share of the ferroelectric phase in the sample gradually increases during the phase transition via intermediate state, whereas the share of the other phase reduces. An antiferroelectric state is piezoelectrically inactive and a ferroelectric state is piezoelectrically active. Due to this circumstance in the case of the phase transition from the antiferroelectric into the ferroelectric state induced by an electric field and realized via intermediate state one may expect a gradual change of piezoelectric properties of the substance.

Here we want to present some experimental results that confirm the above statement. At first we consider conditions for induction of the intermediate state by an electric field and the behavior of the piezoelectric characteristics of materials in the process of induction for several series of the PLZT solid solution with the following compositions: 7.5/100-Y/Y, 8.25/100-Y/Y, 8.75/100-Y/Y. Detailed phase diagrams for solid solutions under investigation are presented in Fig.4.11.

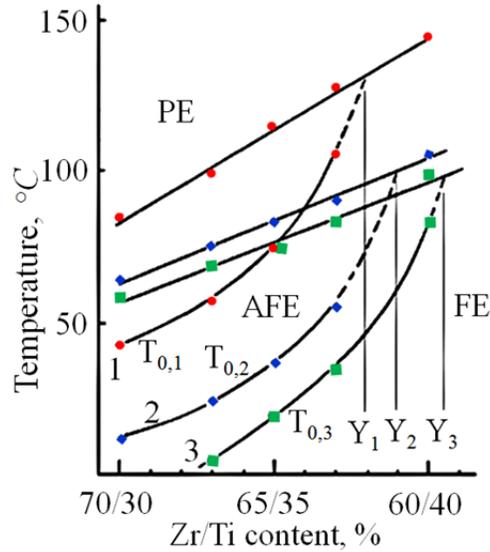

Fig.4.11. Phase diagrams for the X/100–Y/Y PLZT series of solid solutions. La-content (X), %: 1 – 7.50, 2 – 8.25, 3 – 8.75.

The vertical lines $Y_i$ ($i$ = 1, 2, 3) are the boundaries of the regions of spontaneous ferroelectric state (for each series of solid solutions, respectively). For the solid solutions with Ti concentrations located to the right of the vertical lines $Y_i$, the low temperature ($T < T_c$) phase is ferroelectric. In the samples under consideration, the antiferroelectric state is more stable at lower concentrations of Ti. One can induce the ferroelectric state by an external electric field in solid solutions with compositions located near the region of the spontaneous ferroelectric state in the phase diagram (see Fig. 4.11) at low temperatures $T < T_{0,i}(Y)$. This induced ferroelectric state will be preserved under heating until the temperature reaches the characteristic value $T_{0,i}(Y)$. Since the "Ti-composition-temperature" phase diagrams for all above-mentioned series of solid solutions are physically similar in what follows, we present the main results for the solid solutions with composition 8.25/100-Y/Y.

The dependence of polarization on the electric field (the samples were previously annealed at 600 $^oC$) at room temperature is given in the Fig.4.12a. The electric field dependences of the constant $C_f = rf_r$ (here $f_r$ is the frequency of the first radial resonance, and $r$ is the radius of the sample) for the first harmonic of radial oscillations are shown in Fig.4.12b. In contrast to $P(E)$, the dependences $C_f(E)$ are nonmonotonic, therefore, they allow to determine the critical fields of phase transition more precisely.

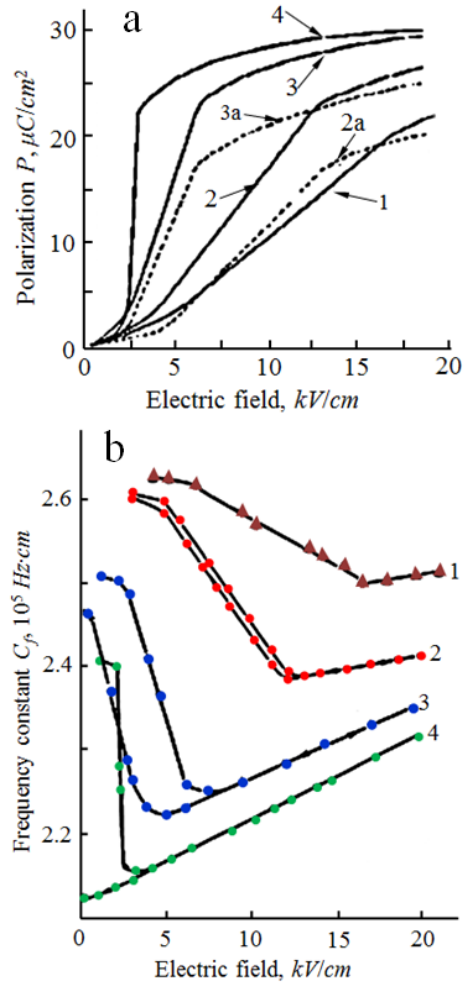

Fig.4.12. Polarization vs. electric field (a) and frequency constant vs. field $C_f(E)$ (b) for the 8.25/100-Y/Y PLZT solid solutions. Zr/Ti content: 1 – 72/28; 2 – 70/30; 3 - 67/33; 4 – 65/35.

As seen from the dependencies in Fig 4.12a and 4.12b, there are three intervals of the electric field intensity, where the properties of solid solutions noticeably differ. Within the interval $0 < E < 4.0$ $kV/cm$ the antiferroelectric state is preserved. At high electric fields ($E > 12.5$ $kV/cm$) the ferroelectric state takes place in the sample. The interval of electric field intensity from 4.0 to 12.5 $kV/cm$ ($E_1 \leq E \leq E_2$, see eq. (4.22) is the region of the intermediate state, and a smooth transformation between the phases participating in the phase transition happens in this range of fields. As seen from Fig.4.12, within this field interval the polarization linearly increases according to expression (4.21). Inside this field interval the displacement of the boundaries between domains take place. The internal state of the domains remains unchanged.

It should be noted that the dependence $P(E)$ is not saturated even at high-intensity fields (up to 20 $kV/cm$ and more). This is connected with the fact that the antiferroelectric domains are still preserved in the volume of the sample after the induction the ferroelectric phase. Such situation was observed in experiments on light scattering in the solid solutions with the composition close to that of the 8.25/67/33 PLZT and belonging to the same region of the "Ti-content-temperature" phase diagram [25, 26]. The increase of polarization at $E > 12.5$ *kV/cm* is connected with the change of the internal state of the preserved antiferroelectric phase domains from antiferroelectric to ferroelectric. This process does not seem to be complete, since there is still no saturation on the curves 2a and 3a in Fig.4.12a which show the sample polarization after the subtraction of $P_i = \varepsilon E$. Completing the discussion of behavior of the 8.25/63/37 PLZT solid solution, we note that the processes caused by the electric field are characterized by a weak hysteresis.

The increase of the titanium content in solid solutions leads to the shift of the *P(E)* and *C(E)* dependences towards weaker fields (see Fig.4.12). Moreover, the interval of the linear increase of the polarization becomes noticeably narrower. This fact is easily explained in terms of the model of the intermediate state considered in [4].

The results discussed above allowed deriving the "Composition-electric field" phase diagrams. They are presented in Fig.4.13 for the 8.25/100-Y/Y and 8.75/100-Y/Y PLZT solid solutions.

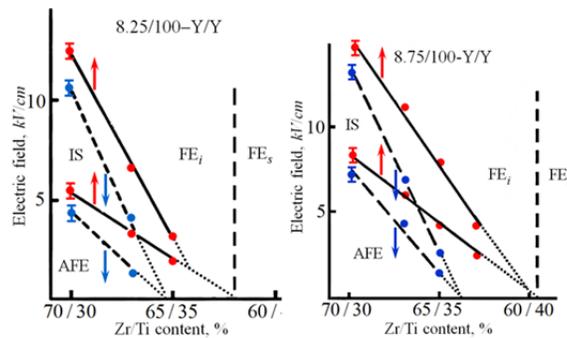

Fig.4.13. "Composition-electric field" phase diagrams for the 8.25/100-Y/Y (a) and the 8.75/100-Y/Y (b) PLZT series of solid solutions.

The intermediate state region (denoted by IS) exists in these diagrams. The denotations $FE_i$ and $FE_s$ are for the induced and spontaneous ferroelectric phases, respectively. In this region the share of the ferroelectric phase increases linearly with the increase of an external DC field. The increase of the share of polar phase leads to changes in the piezoelectric characteristics. It is clearly seen from results of experiments presented in Fig.4.14.

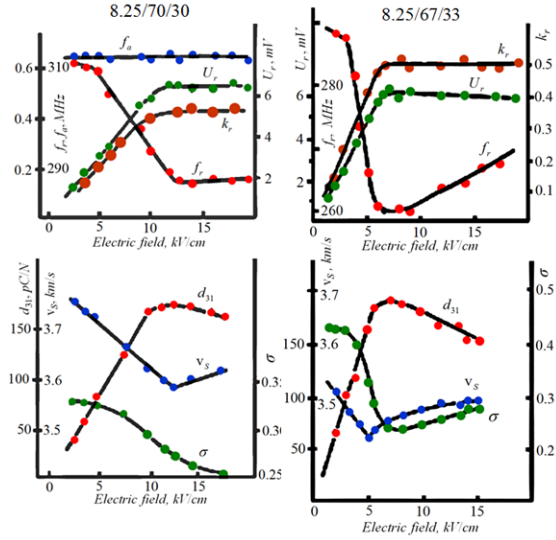

Fig.4.14. Influence of the electric field intensity on the piezoelectric and resonance (radial resonance) parameters of the PLZT series of solid solutions. On the left: the 8.25/70/30 PLZT, on the right: the 8.25/67/33 PLZT.

Based on our experimental results we can conclude that the piezoelectric properties of resonators can be effectively controlled by the external electric field. These results, as well as the "Composition-electric field" phase diagrams (Fig.4.13), directly point to the fact that the interval of effective control of the piezoelectric properties corresponds to the interval of the existence of the intermediate state. The interval of the intermediate state in ferroelectric with coexisting ferroelectric and antiferroelectric phases can be manipulated by proper choice of the ion substitutions as well as solid solutions compositions.

**4.5.** *Materials with negative refractive index for optical range of spectrum*

The possibility of existence of materials with the negative refractive index for electromagnetic radiation and the analysis of the core set of their physical properties was first presented in [89]. Materials with the negative refractive index have been hypothetical for a long period until the first experimental confirmations of hypothesis expressed in [89] appeared in [90, 91]. Since that time researches have taken an active interest in materials of this kind and the number of publications on this topic constantly increases. Comprehensive reviews [92-95] of different physical properties as well as applications of negative refractive index materials have touched practically all aspects of this interesting part of material science existing up to date.

First materials possessing the negative refractive index only in the high frequency region of electromagnetic radiation spectrum represented a composite construction of complex-shape elements (for example, a periodic array of split-ring resonators with wires placed uniformly between the split rings and so on) [90, 91, 96]. It is practically impossible to manufacture similar design for the optical range of spectrum.

The metamaterials with the negative refractive index in the near infrared and optical ranges of spectrum were the metal-dielectric composites [97-99]. The technological approach to manufacturing of composites mentioned in [97-99] is quite laborious and is difficult to reproduce. It has to be noted that the technological process for manufacturing of these composites stays the same even now.

We have developed a simple and reproducible method for fabrication of composite structures with periodic arrangement of high conductivity inclusions in dielectric matrix. These inclusions have sizes in the range of 8 to 12 *nm* with period of arrangement of 0.2 – 2.0 *μm*. The method is based on the process of decomposition of the solid solution in the vicinity of the interphase boundaries between domains of the coexisting ferroelectric (FE) and antiferroelectric (AFE) phases [16, 50, 51].

Lead zirconate titanate based solid solutions are the most studied substances in which the ferroelectric and antiferroelectric ordering takes place. These materials are characterized by a small difference in free energies of the ferroelectric and antiferroelectric states in the solid solution at specified concentrations of components ($PbZrO_3$ and $PbTiO_3$). Domains of the above-mentioned phases coexist in the sample volume of solid solutions belonging to this interval of specific compositions [13, 24, 100, 101]. The metastable phase domains have the shape of cylindrical domains imbedded into the stable phase matrix [13, 24] in the thin crystals or the shape of ellipsoids of revolution close to spheres in the bulk crystals [25]. Such structure of coexisting domains is realized in the crystallites that are not subjected to any external strains. Under the lateral strains the cylindrical domain structure is transformed into the stripe-domain one.

The relative stability of the ferroelectric and antiferroelectric phases depends on the position of the solid solution in the diagram of its phase states (on the relation between the concentrations of components of the solid solution). One can change the relative stability of phases and along with it the volume share of phases in the sample changing the position of solid solution in this diagram. The sizes of the metastable phase domains and their density (the period of the domain structure) change when the share of this phase in the solid solution varies. The density and sizes of the domain structure can be also changed by means of variation of an external electric field intensity as well as mechanical stress (pressure, uniaxial or biaxial compressions or tensile stress). These external influences also change the relative stability of the ferroelectric and antiferroelectric phases.

Conductivity of the majority of oxides with perovskite crystal structure (the PZT-based solid solutions are among them) can be varied by means of ion substitutions in lattice sites in a very wide range from pure dielectric state (with resistivities of the order of $10^{14} – 10^{16}$ $Ω·cm$) to the conducting state (with resistivities of the order of 100 $mΩ·cm$).

Interphase boundaries that separate the adjacent ferroelectric and antiferroelectric domains are characterized by continuous conjugation of crystal planes [13, 24, 45] (without discontinuities of crystal planes and dislocations at the interphase boundaries). Such coherent character has to be accompanied by an increase of elastic energy of the crystal lattice along these boundaries.

As we already discussed the equivalent positions of the crystal lattice in PZT-based solid solutions are occupied by ions with different sizes and/or different electric charges. In the bulk of each domain (away from the domain boundaries) the net force acting on each of these ions is zero. In the vicinity of the interphase boundaries the balance of forces is disturbed, and as a result the "large" ions are driven out into the domains with the larger configuration volume and correspondingly with the larger distances between atomic planes. At the same time the "small" ions are driven out into the domains with smaller distances between atomic planes. Such processes are accompanied the competition of a reduction in the elastic energy along the interphase boundaries on the one hand, and an increase of the energy caused by the deviation of the solid solution composition from the equilibrium composition on the other hand. The processes described above will be finished when the structure of the new interphase boundaries will correspond to the minimum of energy. These new "dressed" interphase boundaries are the interphase boundaries that appear after the redistribution of ions already took place.

The *A*- and *B*-positions of the perovskite crystal lattice of the solid solutions are occupied by ions with different ionic charges ($Pb^{2+}$, $La^{3+}$, $Li^+$ in *A*-sites and $Zr^{4+}$, $Ti^{4+}$, $Nb^{5+}$, $Mg^{2+}$ in *B*-sites). Because of this the local decomposition of the solid solution along the interphase boundaries can be accompanied by the local disturbance of electro-neutrality. The duration of the decomposition process (establishment of the equilibrium heterogeneous composite structure) has the range from several hours to several tens of hours for different solid solutions [13, 14]. That is why one can effectively control the process of formation of the composite structure.

Segregates precipitated along the interphase boundaries are also PZT-base solid solutions [15] but their chemical composition is slightly different from the maternal solid solution. We select the chemical composition of the maternal solid solution and carry out the decomposition in such a way that segregates can possess a diverse set of physical properties (by means of control of the chemical composition of segregates). These segregates can be magnetic, dielectric or conducting.

Powder samples of the PZT-based solid solutions were manufactured by the standard ceramic synthesis process using the co-precipitation of the components from the mixture of aqueous solutions

Samples for optical studies were manufactured by hot pressing method (at a pressure of 30 MPa) at $1250°C$ during 8 hours. Lamellae with the thickness of 0.3 *cm* were cut from the sintered bars. These lamellae were grinded and polished. Before polishing the grinded lamellas were annealed at $1200°C$ in the presence of $PbZrO_3$ filling during one hour and after that at $1100°C$ in oxygen enriched atmosphere during one hour. After the polishing the lamellae were subjected to the second annealing at $850°C$ in oxygen enriched atmosphere during two hours. The wavelength dependences of the light transmission coefficient were measure using Hitachi U-4000 Spectrophotometer.

Initial PZT-based solid solutions have been manufactured with the resistivity of the order of $10^{14}$ Ω·*cm* and the compositions corresponding to the inhomogeneous state of coexisting domains of the ferroelectric and antiferroelectric phases in the volume of the sample. Decomposition of solid solution was performed in such a way that segregates in the vicinity of interphase boundaries had high (close to metallic) conductivity.

The dependences of transparency of one of such materials as a function of the wave length are presented in Fig.4.15 as an example.

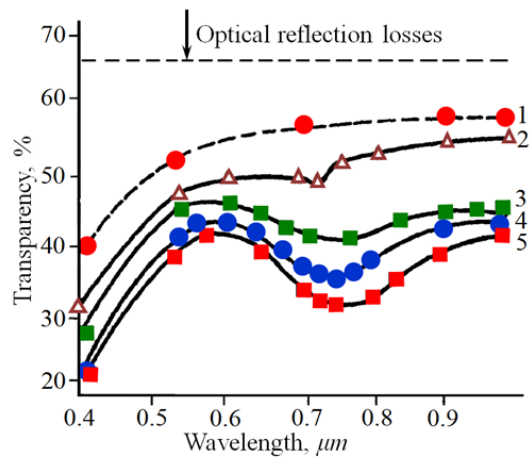

Fig.4.15. Dependence of the light transmission coefficient on the wave length for the transparent composite material (the dielectric matrix containing conductive metallic segregates). The increasing numbers near curves correspond to the materials with increasing value of conductivity of the segregates.

The dips in the curves correspond to the presence of the negative refraction regime. The modulation depth can be controlled by changing the conductivity of segregates. It is also possible to select the optic wave length range by changing the position of the solid solution in the "composition-temperature" phase diagram of the substance. Another possibility is by changing the position of the solid solution in the "electric field-temperature" phase diagram by means of varying the potential difference between the element's electrodes.

However the most interesting results were obtained using thin film structures. The change of the region of negative refraction in this case can be achieved both by the application of the electric field to the film substrate, if the ferroelectric crystal is chosen as a substrate material, and by the flexural strains of the substrate. The modulation of transmitted light has been observed in both these situations.

Based on our results on use of the controlled decomposition of solid solutions for manufacturing of composite materials one can relatively easy develop substances and device components with controlled optical characteristics.